\documentclass[conference]{IEEEtran}
\usepackage{subfigure}
\newcommand{\tabincell}[2]{\begin{tabular}{@{}#1@{}}#2\end{tabular}} 
\usepackage{booktabs}
\usepackage{graphicx}
\usepackage{epstopdf}
\usepackage{tipa}
\usepackage{caption}
\usepackage{floatrow}  
\usepackage{fancyhdr}
\newfloatcommand{capbtabbox}{table}[][\FBwidth]  
\usepackage{amssymb}
\usepackage{bbding}
\newcommand{\ours} {\textsl{GhostTalk}\xspace}
\newcommand{\ourss} {\textsl{GhostTalk-SC}\xspace}

\newcommand{\rev}[1]{{\color{black} #1}}

\usepackage{pifont}
\usepackage{cite}
\usepackage{amsmath,amssymb,amsfonts}
\usepackage{algorithmic}
\usepackage{textcomp}
\usepackage{xcolor}
\usepackage{xspace}
\usepackage{multirow}
\usepackage{url}

\begin{document}
%
\title{GhostTalk: Interactive Attack on Smartphone Voice System Through Power Line}


\author{\IEEEauthorblockN{
Yuanda Wang,   
Hanqing Guo,   
Qiben Yan
}                                     
 \IEEEauthorblockA{
SEIT Lab, Computer Science \& Engineering, Michigan State University\\ \{wangy208, guohanqi, qyan\}@msu.edu 
 }

 }

\maketitle
\thispagestyle{fancy}            
\fancyhead{}                     
\chead{------------------------------The Network and Distributed System Security (NDSS) Symposium 2022------------------------------}                
\cfoot{\quad}                    
\renewcommand{\headrulewidth}{0pt}      
\renewcommand{\footrulewidth}{0pt}

         
\begin{abstract}
Inaudible voice command injection is one of the most threatening attacks towards voice assistants. 
Existing attacks aim at injecting the attack signals over the air, but they require the access to the authorized user's voice for activating the voice assistants. 
Moreover, the effectiveness of the attacks can be greatly deteriorated in a noisy environment. In this paper, we explore a new type of channel, the power line side-channel, to launch the inaudible voice command injection.
By injecting the audio signals over the power line through a modified charging cable, the attack becomes more resilient against various environmental factors and liveness detection models. 
Meanwhile, the smartphone audio output can be eavesdropped through the modified cable, enabling a highly-interactive attack.

To exploit the power line side-channel, we present \ours, a new hidden voice attack that is capable of injecting and eavesdropping simultaneously. 
Via a quick modification of the power bank cables, the attackers could launch interactive attacks by remotely making a phone call or capturing private information from the voice assistants.
\ours overcomes the challenge of bypassing the speaker verification system by stealthily triggering a switch component to simulate the press button on the headphone.  
In case when the smartphones are charged by an unaltered standard cable, we discover that it is possible to recover the audio signal from smartphone loudspeakers by monitoring the charging current on the power line.
To demonstrate the feasibility, we design \ourss, an adaptive eavesdropper system targeting smartphones charged in the public USB ports. 
To correctly recognize the private information in the audio, \ourss carefully extracts audio spectra and integrates a neural network model to classify spoken digits in the speech. 

We launch \ours and \ourss attacks towards 9 main-stream commodity smartphones.
The experimental results prove that \ours can inject unauthorized voice commands to different smartphones with 100\% success rate, and the injected audios can fool human ears and multiple liveness detection models. 
Moreover, \ourss achieves 92\% accuracy on average for recognizing spoken digits on different smartphones, which makes it an easily-deployable but highly-effective attack that could infiltrate sensitive information such as passwords and verification codes. For defense, we provide countermeasure recommendations to defend against this new threat.
\end{abstract}

\section{Introduction}
Smartphone has become an indispensable communication and entertainment tool in everyone's daily life. Many users nowadays spend a substantial amount of time on smartphone apps such as
social networks, mobile games, and live streaming platforms. 
As the technology evolves from text messaging to image and video streaming, the energy consumption of smartphones increases dramatically, creating a pressing demand for large-capacity batteries and fast chargers. 
In recent years, public charger has become a popular utility for travellers in need of charging services, which has grown into a massive billion-dollar market~\cite{chargingstat}. 


Hundreds of millions of users have been using charging stations and power banks all over the world~\cite{user_develop}. The mainstream charging stations can generally be classified into two different types: shared power bank and public charging port.
A shared power bank usually offers different charging cables for different smartphones.
The users typically scan a QR code using their smartphones before renting these power banks, such that they can pay the bill based on the usage time~\cite{guaishouchongdian}. 
Meanwhile, the public charging port, e.g., a USB port, allows the users to charge their smartphones through the port. 
These public charging ports are widely deployed in public spaces, such as shopping malls, hotels, and airports~\cite{publicchargingstation}.

However, even though these charging stations bring convenience to the smartphone users, the ensued security threats have been rapidly escalating. 
For instance, security researchers have exposed numerous attacks that can sniff data transmission through the charging cable~\cite{neugschwandtner2016transparent}, or disclose sensitive app usage from the power consumption profiles~\cite{chen2017powerful}. 
A recent research demonstrates that, by monitoring the input voltage of the charger, an attacker can even recover the smartphone password~\cite{263834}.

From the attacker's point of view, these charging power sources, including the cables and power supply devices, can be modified~\cite{modifiedcable}, and this further exacerbates the threats.
Moreover, the new generation of smartphones have discarded $3.5$mm headphone jack and integrated the headphone audio functions into the charging port~\cite{headphone}.
This innovation revamps the outlook of smartphones, but it imposes new threats to the smartphone audio system when users are charging in public spaces. 
In this research, we find that, after a quick modification of the charging cable, the attackers could have the capability to remotely compromise the smartphone and take control of its voice assistant.

A number of recent studies have demonstrated attacks that compromise voice assistants on smartphones. 
An early study shows that the attackers can compromise speaker verification systems and inject malicious commands into victim smartphones via a replay attack~\cite{wu2015spoofing}.
Yet, the replay of an audible voice command can be easily detected by the nearby victims.
DolphinAttack~\cite{zhang2017dolphinattack} and SurfingAttack~\cite{yan2020surfingattack} both achieve inaudible voice command injection by leveraging the non-linearity of smartphone microphones. 
However, these existing inaudible voice command attacks cannot simultaneously achieve two attack goals, i.e., voice injection and eavesdropping. 

\begin{figure}[t]
    \centering
    \subfigure[In a shopping mall, the victim rents a hacked power bank to charge the smartphone. \ours can remotely compromise the smartphone voice assistant through a modified charging cable.]{\includegraphics[width=8cm]{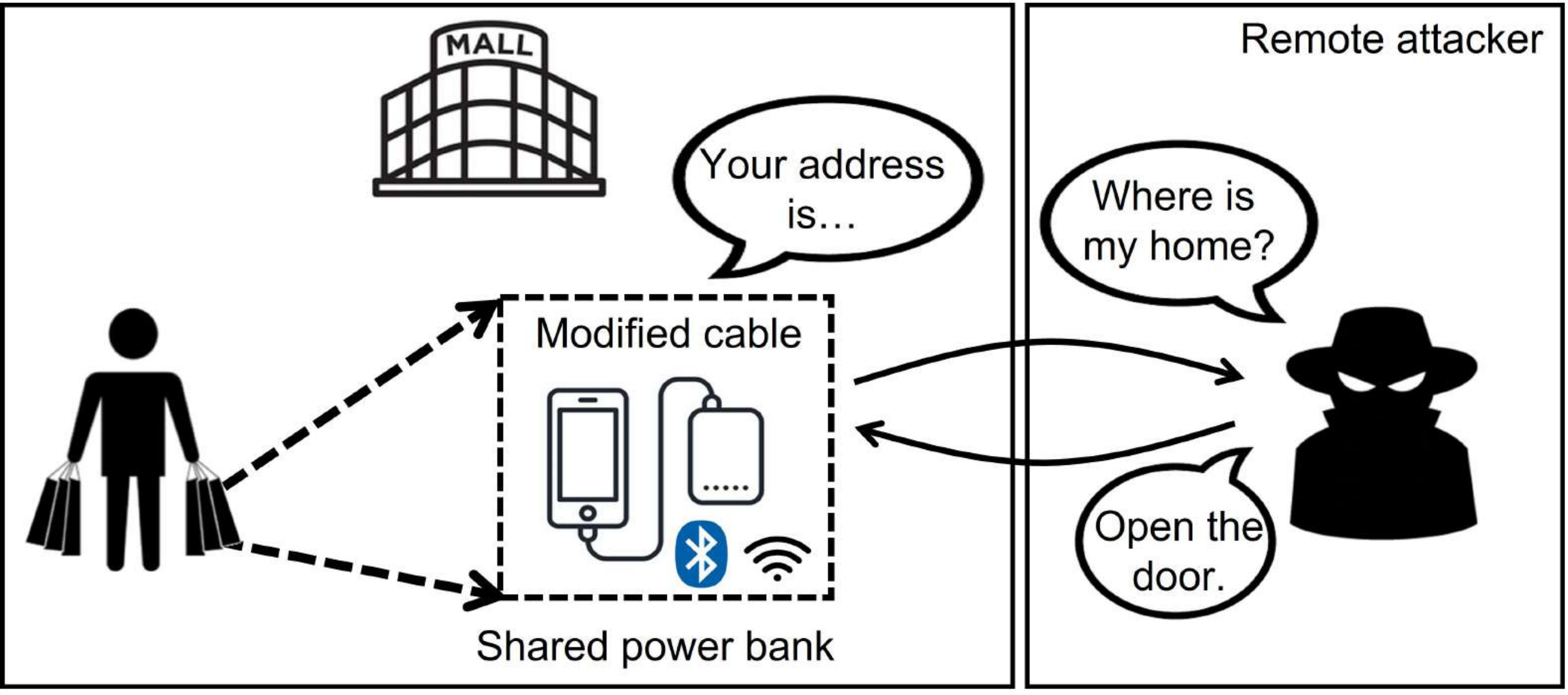}\label{fig:powerbank}}
    \subfigure[The victim answers a phone call when charging his/her phone on a public USB charging port. \ourss eavesdrops private information by analyzing the charging power patterns.]{\includegraphics[width=8cm]{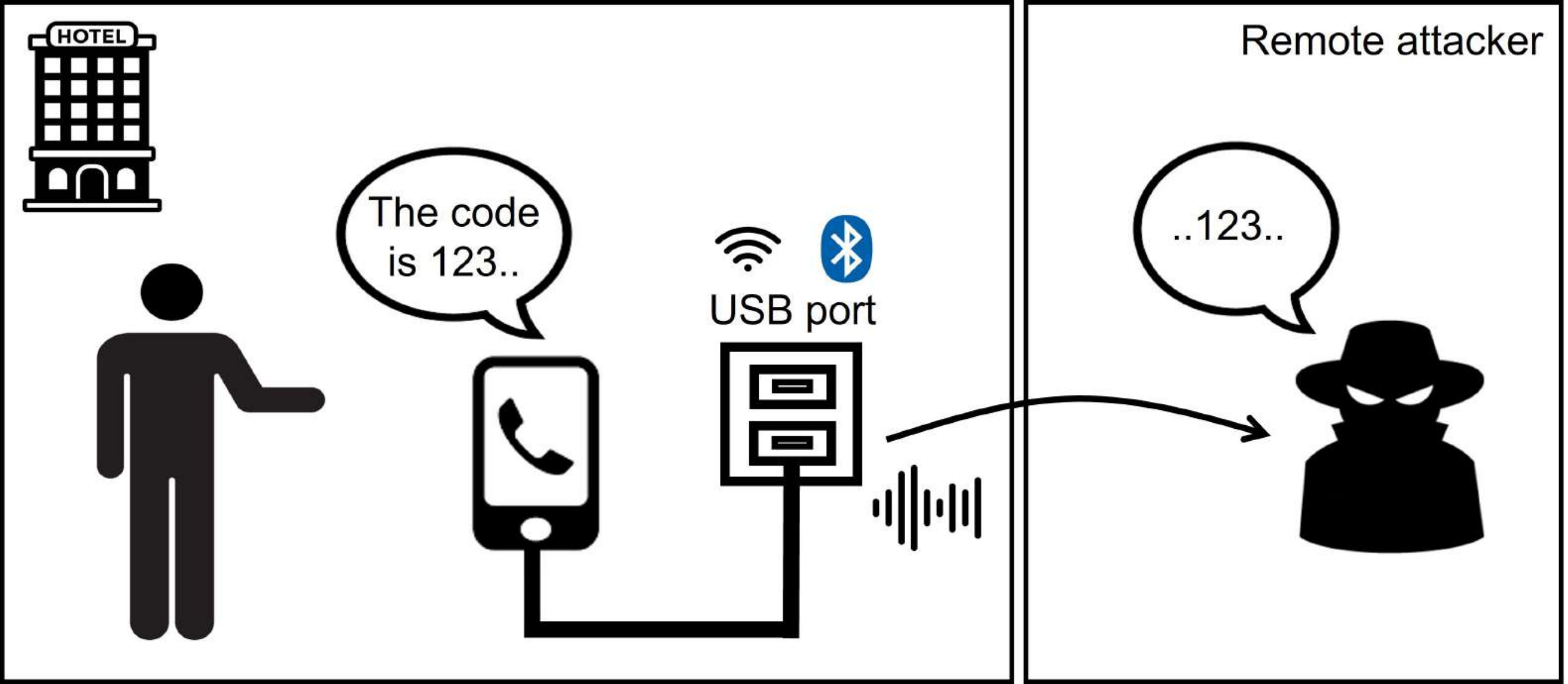}\label{fig: usb}}
    \caption{\ours attacks the smartphones charged by the shared power banks, and \ourss is able to spy private phone conversations when the phone is being charged by the public chargers.}
    \label{fig:power source}
\end{figure}

In other words, the attackers have no way of accessing the responses from the voice assistants. As a result, they are generally incapable of measuring the attack outcome and realizing more complicated attacks, such as ghost phone calls or private information theft.

Moreover, the existing voice injection attacks are susceptible to the environmental noise. 
In order for the attack to succeed, 
the victim device should reside in a quiet environment and the attacker must stay close to it. 
Note that all these attacks suffer from a major shortcoming, i.e., they require the victim's voice to generate specific utterances, like ``Hey Siri" or ``Hello, Google", in order to activate the voice assistant. 
In case when the attacker has no access to the victim's voice, the attack could not be executed. 
Additionally, since these inaudible voice commands are usually transmitted by a loudspeaker, 
they could be effectively detected by the liveness detection modules~\cite{ahmed2020void}.

To further extend the attack scenarios, we introduce \emph{\ours}, a new attack that attempts to compromise the smartphone voice assistants through a power line side-channel.
By modifying the power bank charging cable and manipulating the electric signals in the modified cable, \emph{\ours successfully closes the gap between injection and eavesdropping}, i.e., it not only remotely injects malicious voice commands to the victim smartphone, but it also eavesdrops private information from the voice assistant.
Notably, \ours triggers a switch component to activate the \emph{press} button operation that effectively activates the voice assistant without the requirement of  an authorized speaker's voice.
Fig.~\ref{fig:powerbank} illustrates an attack scenario of \ours: when a user is charging his/her phone with a shared power bank, the attacker can remotely query the user's home address and then unlock the door by interacting with the smartphone's voice assistant.
Compared with the existing work, \ours is the first interactive attack that simultaneously achieves stealthy audio injection and eavesdropping, while remaining resilient against environmental noises and liveness detection systems. 



\rev{In another attack scenario when the users are charging their phones on the public charging ports, they typically insert their own standard charging cables.  
We experimentally observe that the power usage patterns of the phone's loudspeaker could be used as a side-channel for extracting private audio signals.}
Specifically, when the battery is over 95\% charged, the attackers could extract the audio signal by passively monitoring the charging current. 
Based on this observation, we design \emph{\ourss} (i.e., \ours with Standard Cable) to eavesdrop sensitive information from the smartphones charged by standard cables.
During the attack, whenever the victim is playing audio through the smartphone loudspeaker, the adversary could recognize the leaked audio by measuring and analyzing the varying charging power.
\rev{However, the background noise introduced by other smartphone applications has a substantial impact on the perceptibility of the captured audio.}
To overcome this challenge, \ourss denoises the audio by signal processing and leverages deep neural networks (DNNs) to recognize sensitive digits within the conversation. A website is set up (\textbf{\url{https://ghosttalkattack.github.io/}}) to demonstrate the attacks. 

In summary, this paper makes the following contributions:
\begin{itemize}
    \item \ours is the first interactive attack towards smartphone voice assistants over the charging cables. After slight modifications on the charging cable of the shared power banks, \ours can achieve interactive attacks by inaudible audio injection and eavesdropping. 
    In addition, \ours attack requires no prior knowledge about the victim's voice and preserves the resilience against noisy environments and liveness detection models.
    \item We propose \ourss, an eavesdropping attack that captures the audio signals from the power line side-channel. \ourss can successfully extract audio signals from 8 out of 9 tested smartphones through standard charging cables, and can accurately recognize sensitive digit information using a DNN model.
    \item We evaluate \ours and \ourss attack performance with extensive real-world experiments using 9 popular commodity smartphones. The results prove that \ours  achieves inaudible interactive voice injection and eavesdropping attacks on all the victim smartphones. Moreover, \ourss can correctly classify more than 90\% of spoken digits in the leaked audio when the smartphone plays the audio at its highest volume.
\end{itemize}

\section{Background}\label{background}
\subsection{Smartphone Charging Ports}
\begin{figure}
    \centering
    \subfigure[Lightning port circuit.]{\includegraphics[width=3in]{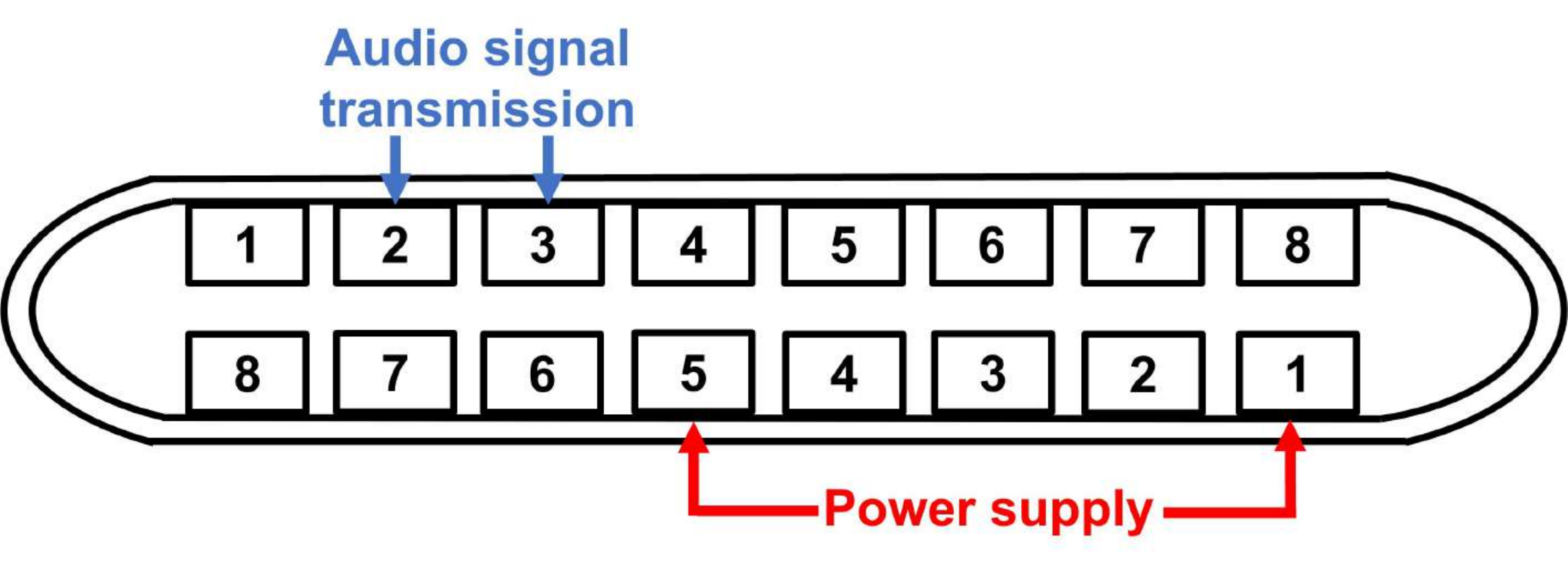}\label{fig:lightening}}
    \subfigure[USB-C port circuit.]{\includegraphics[width=3in]{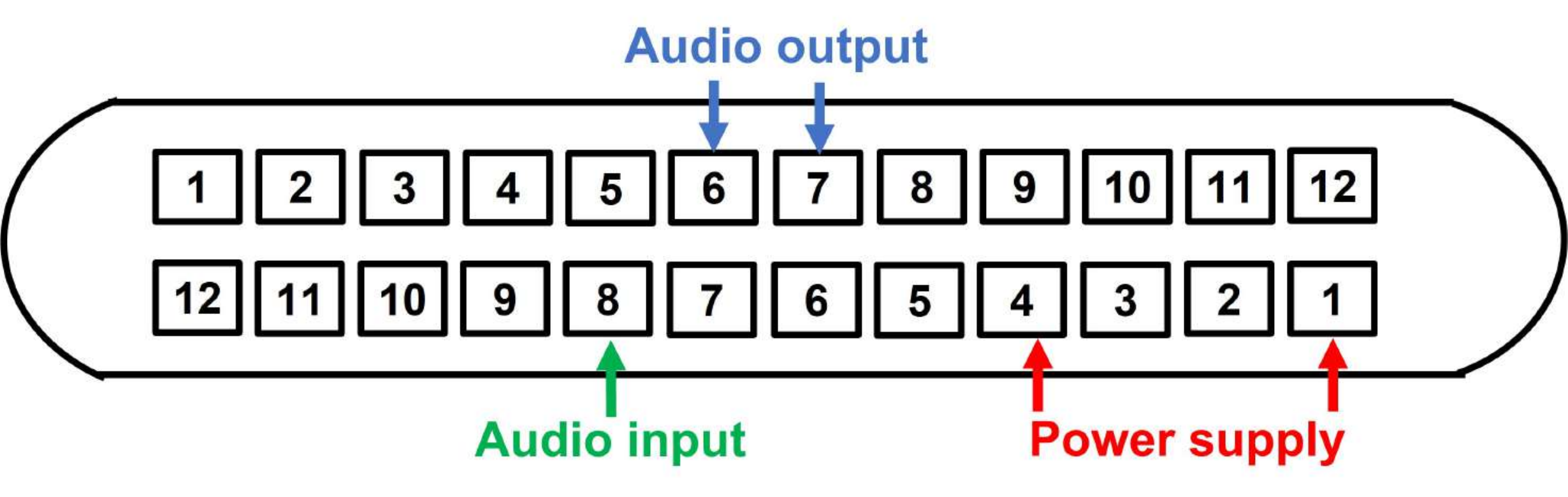}\label{fig: typec}}
    \caption{Two mainstream charging port architectures of smartphones.}
    \label{fig:ports}
    \vspace{-5pt}
\end{figure}
Traditional charging ports on the smartphone possess two main functions: charging and data transmission.
On the new generation of smartphones, the manufacturers are trending towards the complete removal of the headphone jacks, while supporting the audio signal transmission directly over the charging ports. 
Correspondingly, two mainstream charging ports, Lightning port and USB-C port, both support the audio transmission over the charging port.

Fig.~\ref{fig:lightening} illustrates the circuit of Lightning charging ports equipped on iPhones. 
Generally, Lightning port can work under four modes: USB host, USB device, accessories, and power supply.
\rev{The accessories mode supports the  concurrent battery charging and audio transmission. Specifically, pin 2 and pin 3 transceive the audio signals, while pin 1 and pin 5 are responsible for charging the battery.}


The USB-C port widely deployed in Android smartphones is shown in Fig.~\ref{fig: typec}.
When a headphone is plugged in, pin 6 and pin 7 will send the audio signals to the headphone, and pin 8 will receive the input audio signal from the microphone. 
Meanwhile, pin 1 and pin 4 connect to the DC power for charging. 
Therefore, USB-C also simultaneously supports charging and audio signal transmission.
Mainly due to the integrated and versatile features of these ports, the smartphones are threatened by the unauthorized audio injection and eavesdropping attacks as shown in this work.


\subsection{Headphone Circuit}
\begin{figure}
    \centering
    \includegraphics[width=8cm]{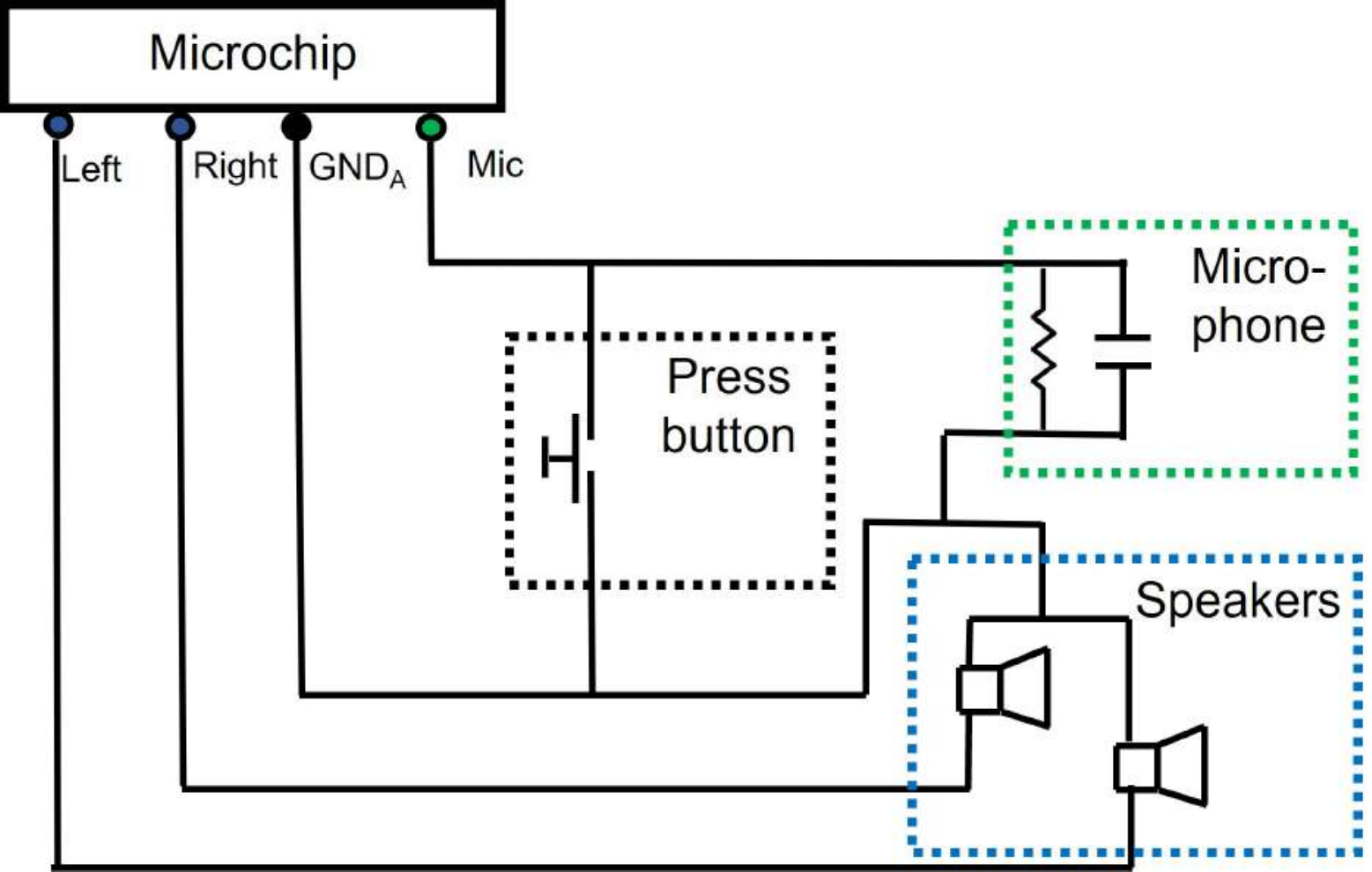}
    \caption{A typical circuit of a wired headphone with microphone component and press button.}
    \label{fig:headphone}
\end{figure}

Fig.~\ref{fig:headphone} displays the circuit of a typical wired headphone with Lightning or USB-C jack. 
In the headphone, 4 wires transceive audio signal from/to the smartphone: left speaker, right speaker, microphone (Mic) and audio ground (GND$_{A}$). 
When the headphone is playing audio, the smartphone outputs digital signals to the charging port, where the microchip digital to analog converter (DAC) converts them into analog voltage signals.
After that, the voltage signals will trigger the change in the current of the headphone speaker coil. 
Such a changing current in turn stimulates the vibration of the speaker membranes to generate the audible sound wave.

Conversely, the sound waves cause membrane vibrations that modify the microphone capacity.
As the voltage on the capacitor is constant, the changing capacity translates into the changing current, which produces the analog signals corresponding to the input audio. 
The microchip analog to digital converter (ADC) will then convert the analog audio signals into digital data, and transmit the data over to the smartphone.

Most of the smartphone headphones have a ``press" button to allow smartphone operations such as making phone calls or controlling music players. 
When the button is clicked, the microphone and audio ground are shorted and the smartphone detects a current impulse from the microphone. 
It is noteworthy that the press button can also activate the voice assistants, and this function is exploited by \ours to enable the hidden activation,
as shown in Section~\ref{systemdesign}.

\rev{
\subsection{Power Line Side-channel}
Li-ion battery is widely deployed on smartphones. 
Generally, the charging process of a Li-ion battery can be divided into three stages~\cite{battery}: (i)
with a low battery status, the charger will offer a constant current to boost the battery voltage;
(ii) during the charging process, the charging current adjusts to keep the charging voltage constant;
(iii) once the battery is fully charged, the charging power is consumed to balance the smartphone power usage.
At the final stage, the charging power is determined by both the smartphone hardware components and the running apps. 

Recent work demonstrates that the charging power 
has a strong correlation with the smartphone apps when the battery state is over 95\%~\cite{10.1145/3460120.3484733}.
As a result, the charging power pattern can reflect the smartphone's working status, thereby opening up a side-channel for the attackers.
For example, the attackers can fingerprint specific websites and apps by recognizing different charging power patterns~\cite{yang2016inferring,chen2017powerful}, 
or even steal the lock-screen password by measuring the charging voltage fluctuation~\cite{263834}. Our work develops new attacks to extract the audio signals from the power-line side channel. 
}

\section{Attack Motivation}\label{motivation}
\subsection{Cable Modification}\label{fakecable}
To implement the attack, the attacker has to modify the standard charging cables in order to support audio signal transmission. 
However, it will be extremely difficult for attackers to add audio functions in a standard cable.
Fortunately, we can use the headphone adapter, whose cable allows concurrent audio signal transmission and charging, which is very popular on the market with a fair price ($\sim$\$10). 

\begin{figure}
    \centering
    \includegraphics[width=8cm]{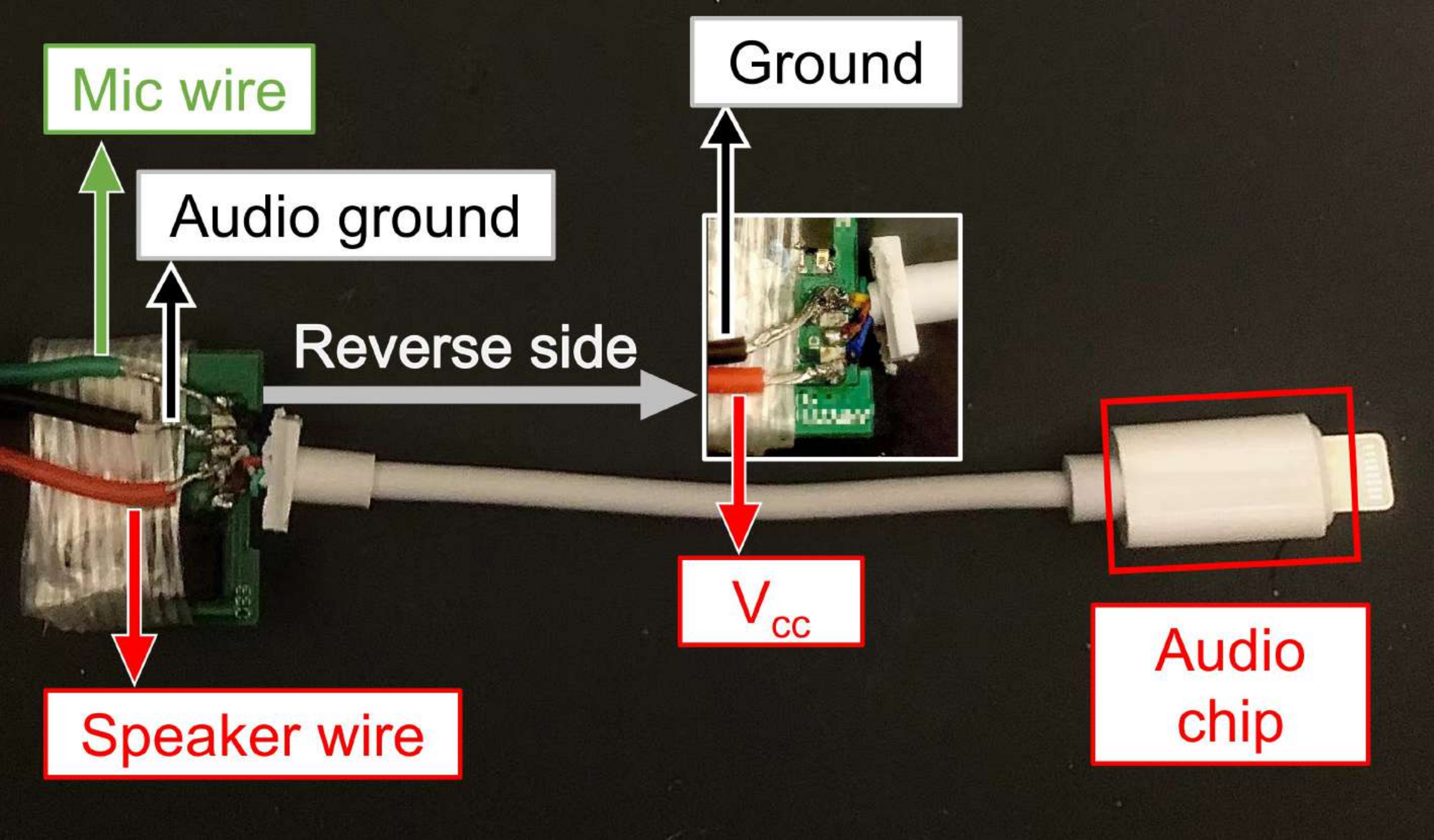}
    \caption{The modified charging cable for \ours attack.}
    \label{fig:adapter}
\end{figure}

Fig.~\ref{fig:adapter} shows a specially designed Lightning adapter cable that enables audio functions and charging. 
By integrating audio functions in the microchip, the cable can encode and decode audio signals. 
Two charging wires, as shown in the middle box of Fig.~\ref{fig:adapter}, charge the smartphone, while the four extra audio wires are used for audio signal transmission (see the left boxes in Fig.~\ref{fig:adapter}). Similar adapter cable exists 
for USB-C. 
The attackers can then replace the standard cables of the shared power bank with such specially designed cables and launch attacks towards the smartphones being charged.

\subsection{Inaudible Audio Injection through Charging Cable}\label{injmotivation}

\begin{figure}
    \centering
    \subfigure[The voltage measurement result between microphone wire and audio ground.]{\includegraphics[width=0.47\columnwidth]{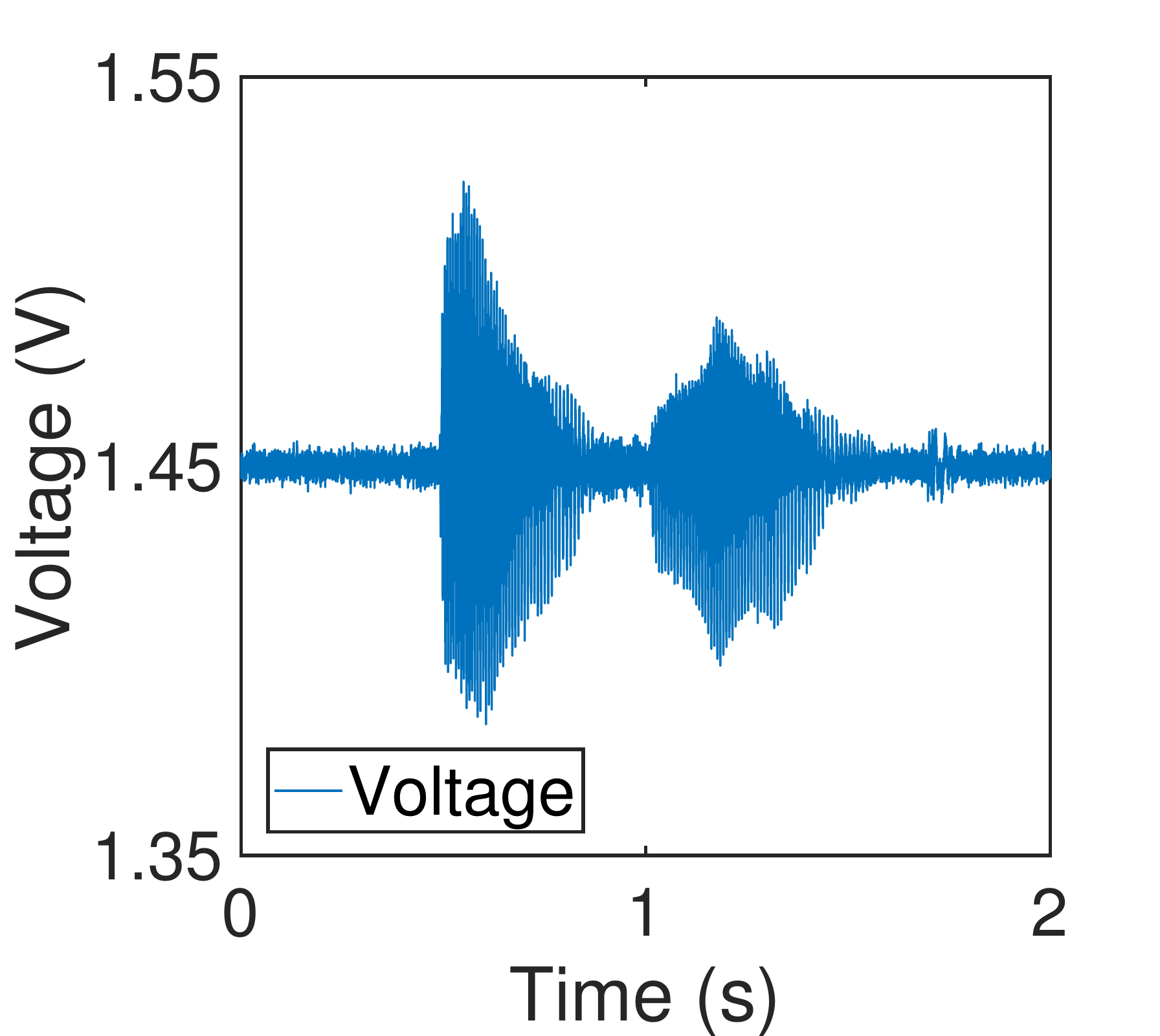}\label{fig: inj_vol}}
    \centering
    \subfigure[Injected audio waveform recorded by the victim smartphone.]{\includegraphics[width=0.47\columnwidth]{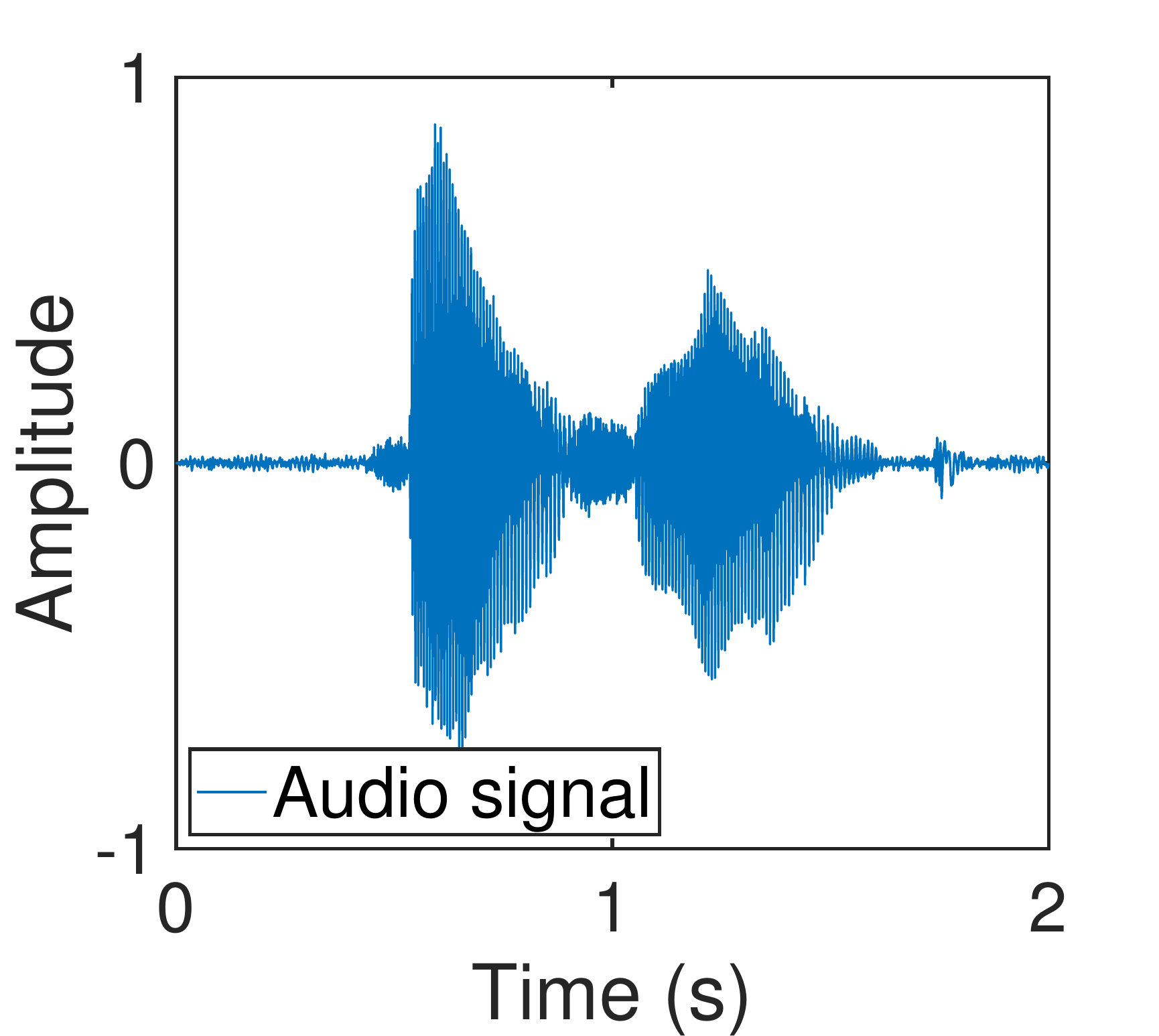}\label{fig:inj_raw}}
    \caption{The relationship between the voltage on microphone wire and the signal strength of the corresponding recorded audio.}
    \label{fig:mic_wave and voltage}
    \vspace{-5pt}
\end{figure}
As illustrated in Section~\ref{background}, the audio signals over the charging port are essentially represented by the changing current.
Therefore, if the attackers can manipulate the current through the audio wires, they can inject the inaudible audio signals into the smartphone.

To verify the feasibility of audio signal injection through a charging cable, we add modulated voltage signals between the microphone and audio ground to change the current in the microphone.
Specifically, we add an extra DC offset ($\sim$1.45V) to the modulated signal, and apply it on the microphone wire. 
Then, the victim smartphone, i.e., iPhone X, records the injected audio signal from the Lightning port. 
Meanwhile, we use an ADC board 
to measure the input voltage signal.
\rev{Fig.~\ref{fig: inj_vol} shows the injected voltage waveform of a voice command ``Hey Siri", and the resulting audio waveform is presented in Fig.~\ref{fig:inj_raw}.
Apparently, the shapes of the voltage and audio waveforms resemble each other. 
Hence, the results prove the feasibility of inaudible voice command injection by controlling the voltage input on the microphone wire of a charging cable. 
This phenomenon demonstrates the existence of a charging port backdoor that can be exploited to stealthily attack the voice assistants.}
 



\subsection{Inaudible Audio Eavesdropping though Charging Cable}\label{eavmotivation}

\begin{figure}[htbp]
    \centering
    \subfigure[The original audio waveform played by the victim smartphone.]{\includegraphics[width=0.47\columnwidth]{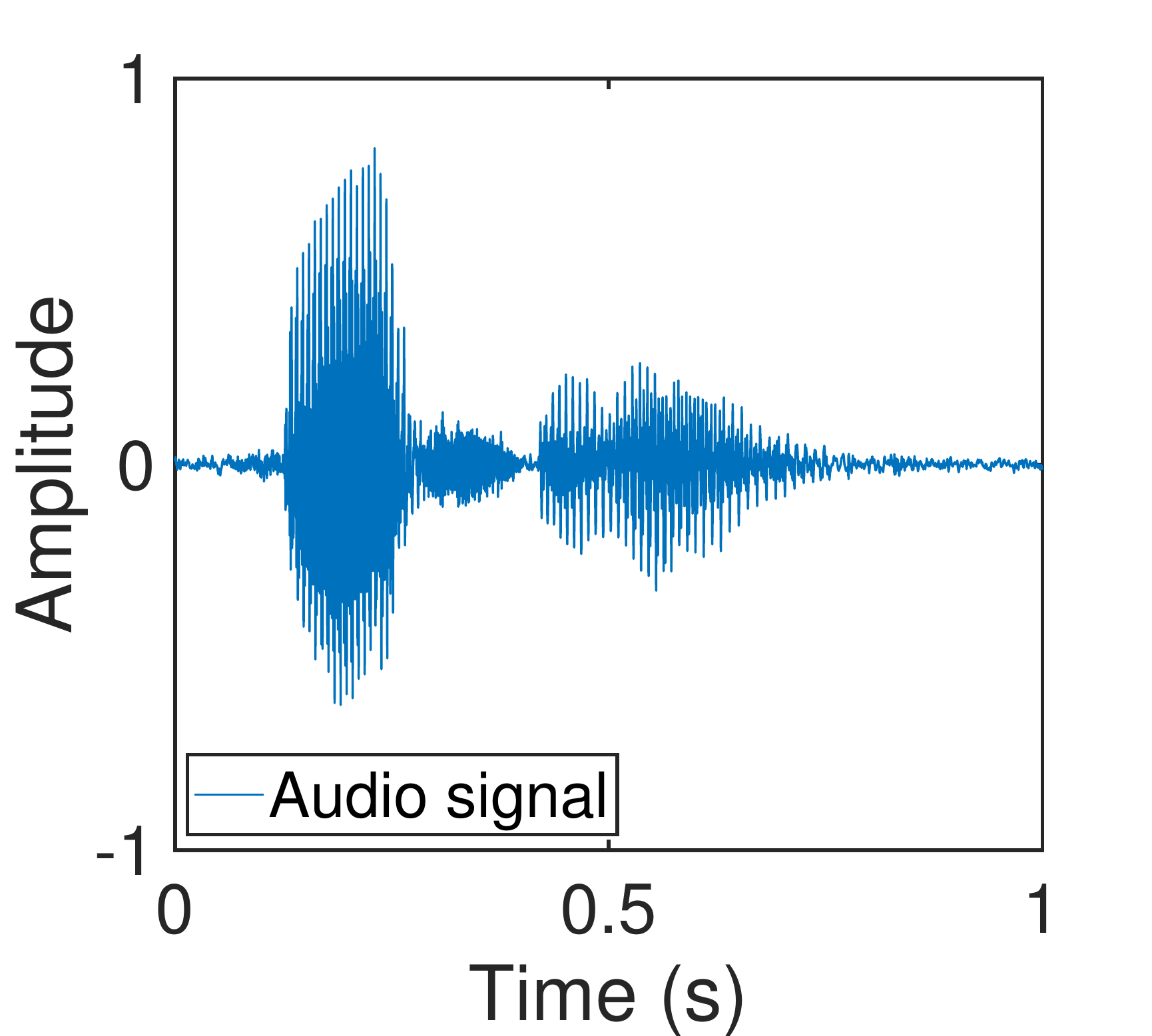}\label{fig:eav_raw}}
    \centering
    \subfigure[The voltage measurement result between speaker wire and audio ground.]{\includegraphics[width=0.47\columnwidth]{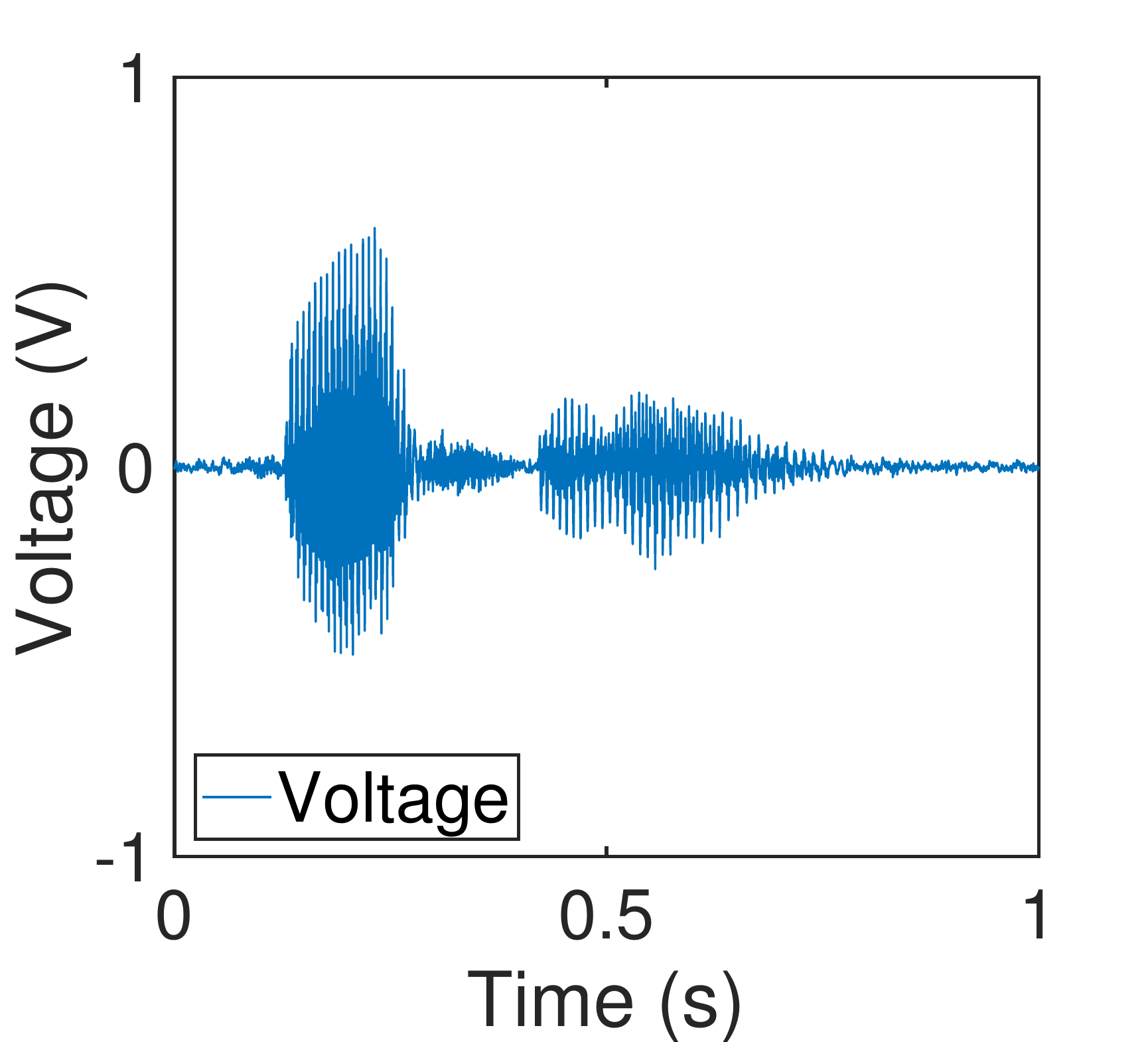}\label{fig: eav_vol}}
    \caption{The relationship between the signal strength of the played audio and the voltage on the speaker wire.}
    \vspace{-5pt}
    \label{fig:speaker_wave and voltage}
\end{figure}
Next, we evaluate the feasibility of eavesdropping by monitoring the voltage signal on the charging cable. 
First, we play a recorded word ``password" on the same iPhone X, and monitor the voltage between the speaker and audio ground wires. 
The original audio waveform is shown in Fig.~\ref{fig:eav_raw}, while the measured voltage waveform is shown in Fig.~\ref{fig: eav_vol}. 
The voltage waveform almost perfectly resembles the audio waveform, which demonstrates that the audio signal can be accurately recovered by voltage measurement. 
\subsection{Audio Eavesdropping through Standard Charging Cable}

\begin{figure}[t]
    \centering
    \includegraphics[width=8cm]{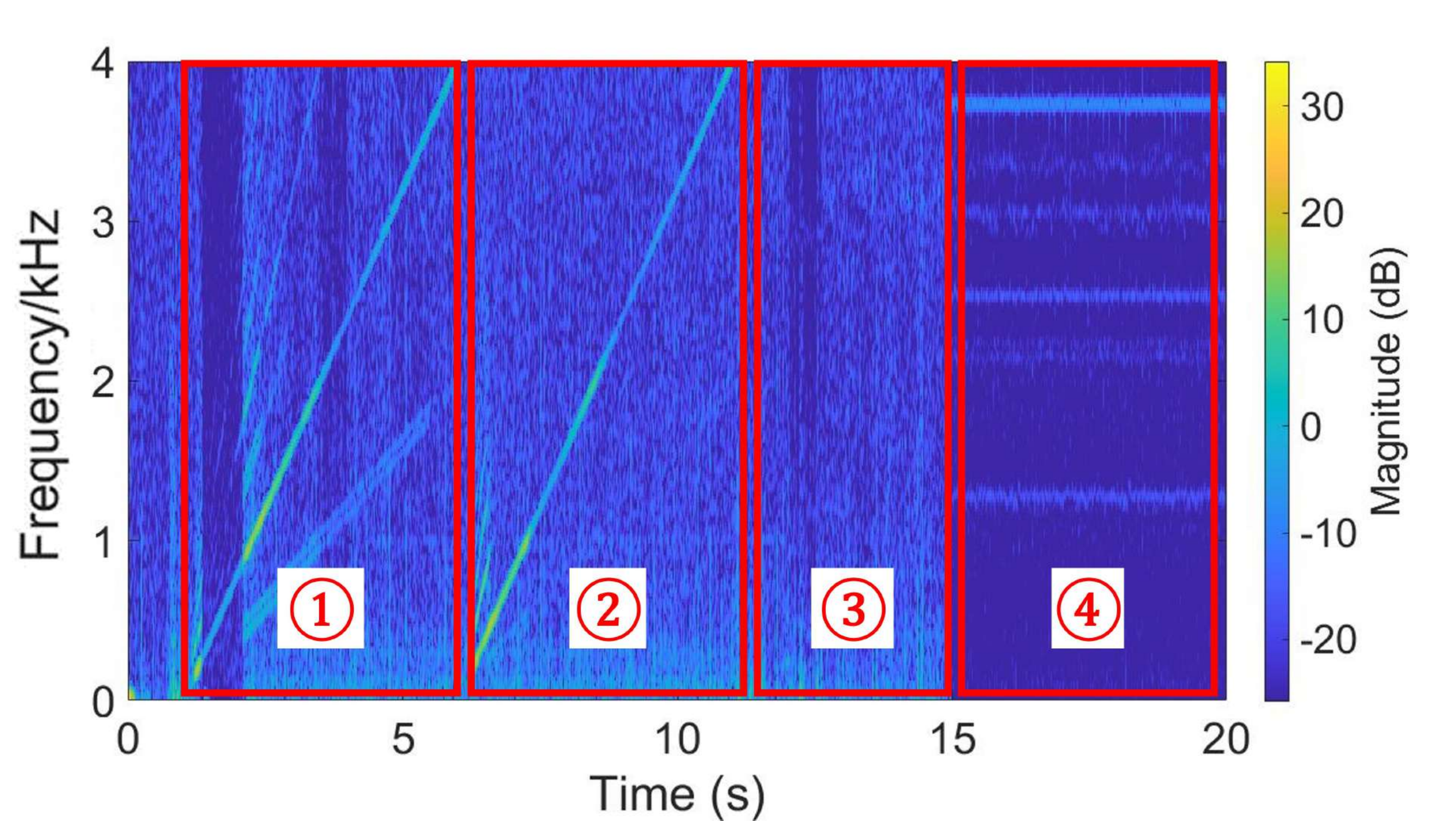}
    \caption{\rev{When the smartphone plays a chirp audio signal via bottom speaker (\ding{192}) and top speaker (\ding{193}), the charging current spectrogram contains the chirp signal patterns. 
    When the smartphone idles, the charging current spectrogram still carries the noise  brought by the smartphone firmware and apps (\ding{194}). The noise disappears once the smartphone is turned off (\ding{195}).}}
    \label{fig: chirp}
\end{figure}

\begin{figure}[t]
    \centering
    \includegraphics[width=8cm]{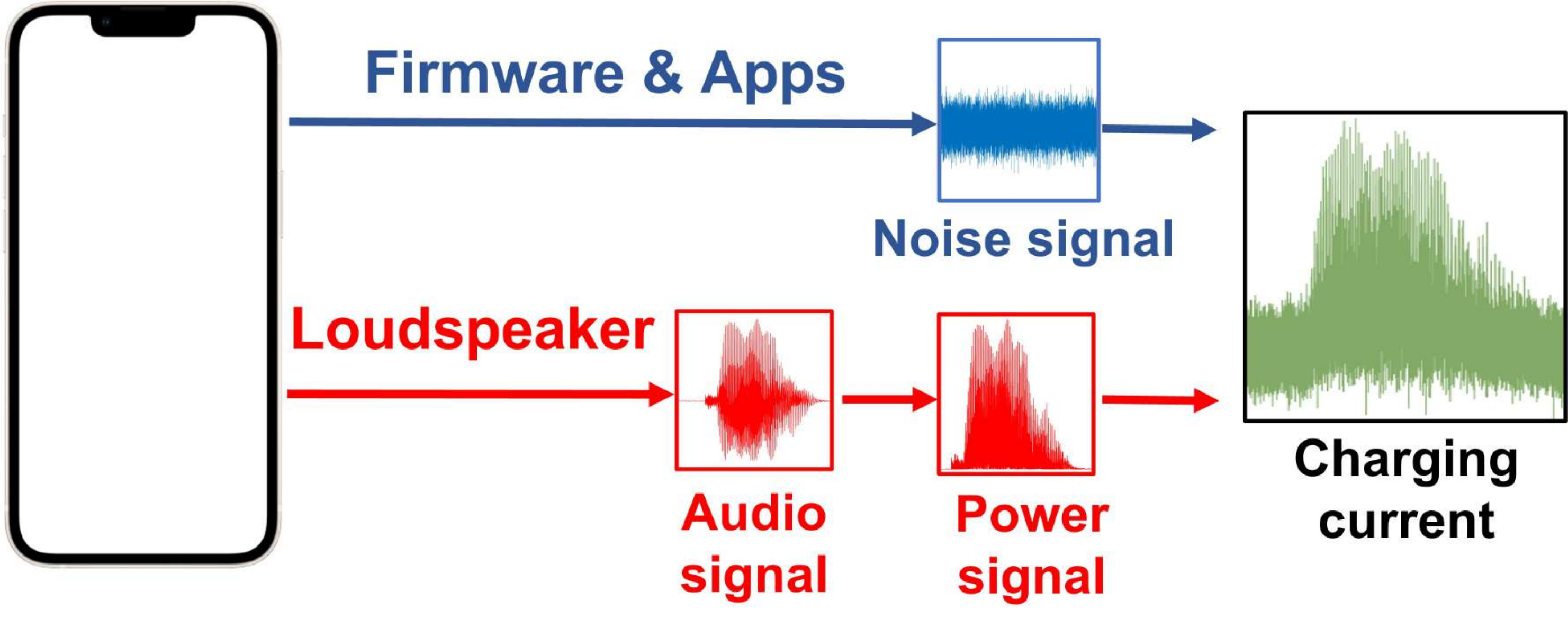}
    \caption{\rev{Power side-channel leaks smartphone audio signals, but they are obfuscated by strong noise.}}
    \label{fig: powersidechannel}
\end{figure}

\rev{
In case when the users plug their own standard charging cables in the public charging ports, 
the attackers could not access and modify these cables. 
However, we find that the charging power line side-channel could still leak the audio signal. 
This power side-channel may be caused by the high power profile of the loudspeaker, or the electromagnetic (EM) field from the loudspeaker that alters the charging current due to the close proximity of the loudspeaker and charging port components. 
We design four experiments to find the root cause of this power line side-channel.
Particularly, we measure the charging current of a fully-charged iPhone X via a shunt resistor.
First, the phone plays a chirp audio (0$\sim$2 kHz) through the left audio channel, 
originating from the bottom loudspeaker. 
Second, the phone plays the same chirp audio through the right audio channel, originating from the top loudspeaker. 
Third, the smartphone idles after playing the audio. 
Finally, we take additional measurement when the smartphone is off.

Fig.~\ref{fig: chirp} presents the charging current spectrogram under different experiments.
The results in \ding{192} and \ding{193} show that the signal strengths of charging current are  exactly the same regardless of the positions of loudspeakers (i.e., top or bottom). Therefore, the power-line side channel is unlikely induced by EM interference, which varies notably across different positions. 
Fig.~\ref{fig: chirp} also shows that the frequency of the charging current signal (0$\sim$4 kHz) doubles that of the audio signal (0$\sim$2 kHz). 
In fact, when the loudspeaker is playing a $k$ Hz audio signal, its power consumption $P_{l}$ can be expressed as:
\begin{equation}
    P_{l}=I^{2}_{l}R=(a\textnormal{cos}(2\pi k t))^{2}R=\frac{a^{2}R}{2}(1+\textnormal{cos}(4\pi k t)),
    \label{powermodel}
\end{equation}
where $a$ is a constant and $R$ is the resistance of the loudspeaker. 
Eq.~(\ref{powermodel}) illustrates that the frequency of power usage doubles that of the audio signal, which perfectly matches with our experimental result. Therefore, we can see that the audio signal patterns in the charging current is brought by the high power profile of the loudspeaker, which far exceeds the idling charging power. 

However, as shown in Fig.~\ref{fig: powersidechannel}, since the smartphone firmware and apps also draw power, the leaked audios would be too noisy to be recognized by human ears.
Therefore, we use a convolutional neural network (CNN) to recognize sensitive information in the speech audio as shown in Section~\ref{systemdesign}.
Also, given that the noise level is associated with the smartphone hardware design, the signal strengths of leaked audios vary significantly for different smartphones. The details are presented in Section~\ref{evaluation}.
}


\section{Threat Model}\label{threatmodel}

\begin{figure*}[htbp]
    \centering
    \subfigure[Private information query.]{\includegraphics[height=3.7cm]{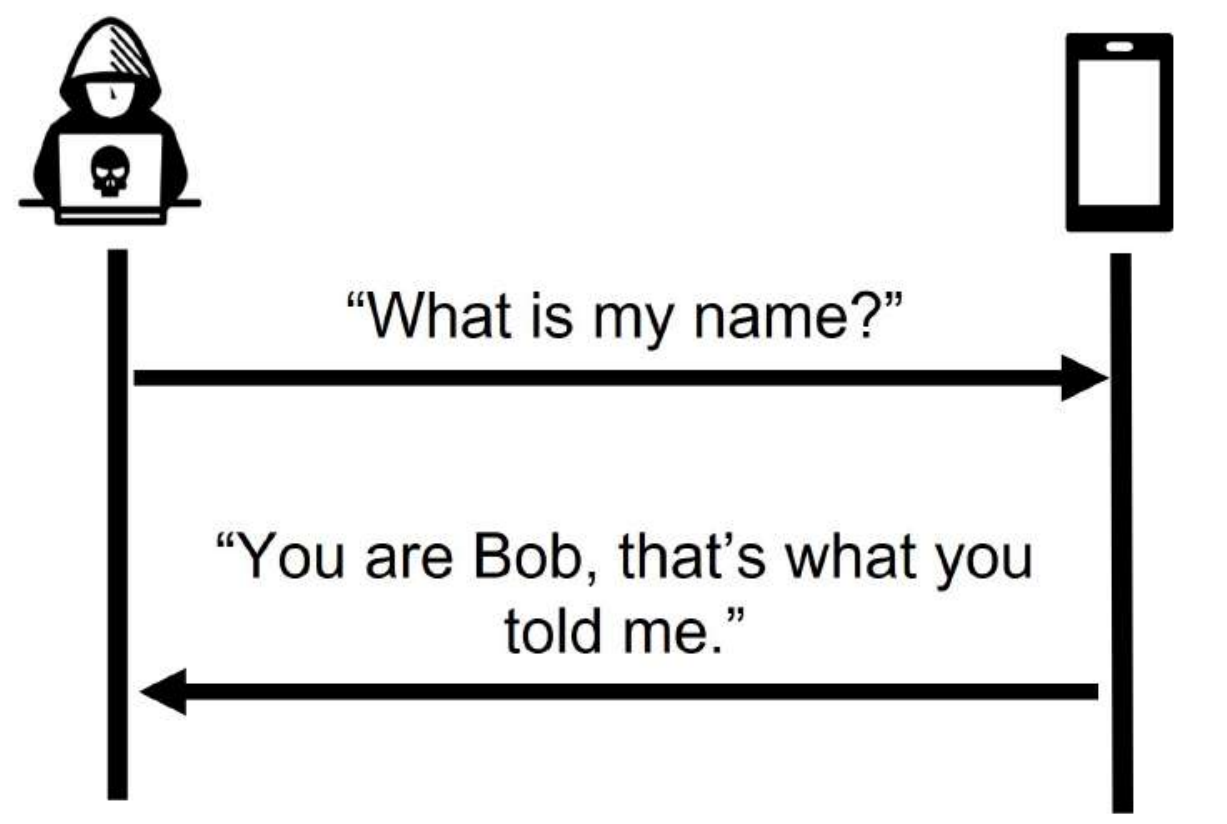}\label{fig: conversation}}
    \subfigure[Ghost phone call.]{\includegraphics[height=3.7cm]{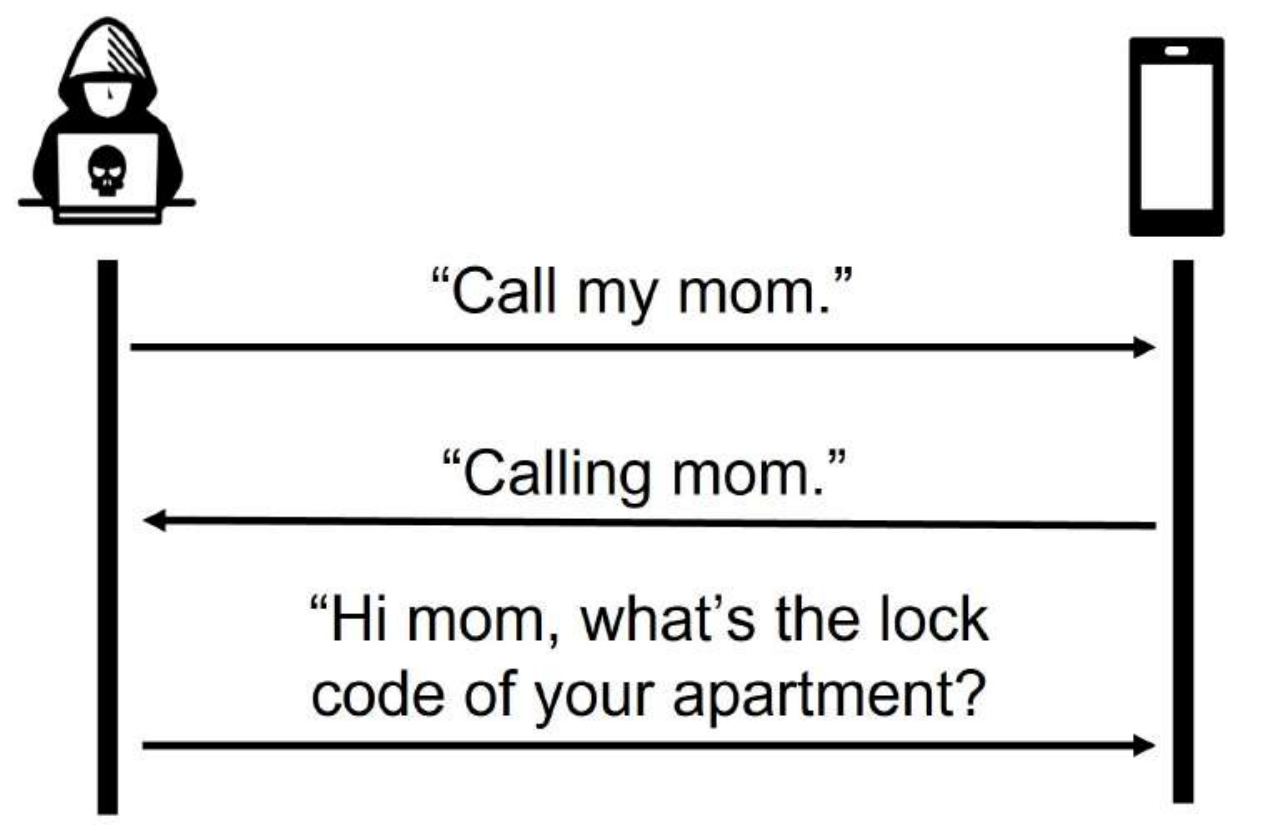}\label{fig: ghost call}}
    \subfigure[Hacking verification code.]{\includegraphics[height=3.7cm]{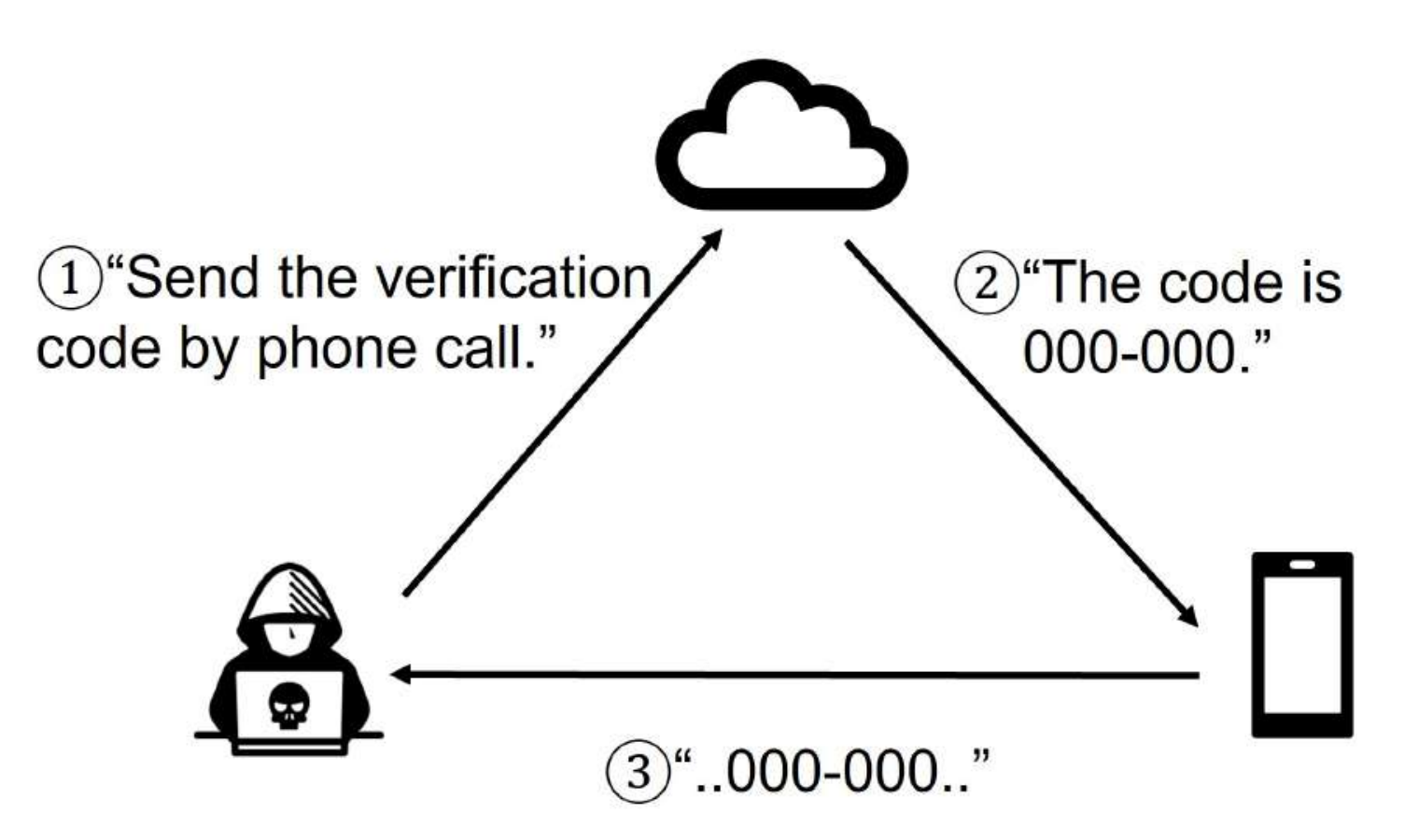}\label{fig: code eavesdropping}}
    \caption{Specific attack scenarios of \ours.}
    \label{fig:attack scenarios}
\end{figure*}
\subsection{Threat Model of \ours}
\ours works when the modification of charging cables of a shared power bank is a viable option. 

\rev{\noindent\textbf{Power Bank Modification.} Since everyone has access to the shared power banks, it is reasonable to assume that the attacker can replace the charging cables of the shared power bank  with specially designed cables, and hide extra hardware in the power banks. The victim rents a (hacked) power bank to charge the smartphone in a public space such as airport, hotel, or shopping mall. 
}

\noindent \textbf{No Owner Interaction.} We assume that the victim users do not keep using their smartphones, when the phones are being charged by the power banks. This is a common assumption taken by almost all the voice command injection attacks~\cite{zhang2017dolphinattack}. 
For example, it is quite normal for people to put their phones in a handbag along with power banks.

\rev{\noindent \textbf{Attack Scenarios.} 
For \ours attack, the attacker does not need to be physically close to the victim device, as the attack device (i.e., modified power cable) contains a WiFi module. By connecting to public WiFi hotspots, the attacker could launch the attack remotely by sending/receiving audio signals from a remote site.
Fig.~\ref{fig:attack scenarios} shows three specific attack scenarios of \ours:
(a) the attackers can query the voice assistant to steal private information, such as the user's name, home address, and phone number; 
(b) after the attackers retrieve the victim identity, they can collect or generate the victim's voice samples by crawling the social media or running speech synthesis, and search for the victim's family and job information on the Internet.
Then, the attackers can launch a ghost phone call by injecting and eavesdropping voice signals
as shown in Fig.~\ref{fig: ghost call};
(c) the attackers can request a voice verification code to be sent to the smartphone, and stealthily eavesdrop it upon the reception.
The verification code can be used to hack into the victim's social media or bank accounts.}

\subsection{Threat Model of \ourss}
Although the \ours attack brings a notable threat, it can only be implemented on smartphones being charged by modified power banks.
To further extend the attack scenario, an alternative attack, \ourss, works without the need of charging cable modification. 

\noindent \textbf{Power Source Modification.} Recent work~\cite{lau2013mactans} shows the feasibility of hacking public USB charging ports by attaching malicious hardware.
For \ourss attack, the attacker can also hide an ADC board in the USB ports of hotels and airports. 
The ADC board will keep monitoring the charging current and send the measurement results to the attacker.

\noindent \textbf{Victim's Behavior.} 
We assume the victim will not immediately stop charging after the battery state exceeds 95\%. \rev{We also assume the victim will raise the loudspeaker volume when interacting with the phones in a hands-free mode, which is very common in our daily life.}

\noindent \textbf{Attack Scenario.} \rev{If the victim keeps charging the smartphone after the battery state reaches 95\%, the audio signal from the loudspeaker becomes extractable by the attacker. Despite the charging state assumption, the attack scenario is still quite realistic. 
As an example scenario, in a public space, a victim plugs the smartphone on a wall-mounted charging port. 
While charging, the victim retrieves verification codes or delivers private information such as credit card information and SSN number over a phone call. It is not unusual that the verification codes and conversations are played aloud by the smartphone speaker. 
\ourss can then eavesdrop the voice verification codes or passwords by recognizing the charging current patterns.}
\section{Attack system design}\label{systemdesign}
\subsection{System Design of \ours}\label{attack_ours}
\begin{figure}
    \centering
    \includegraphics[width=3in]{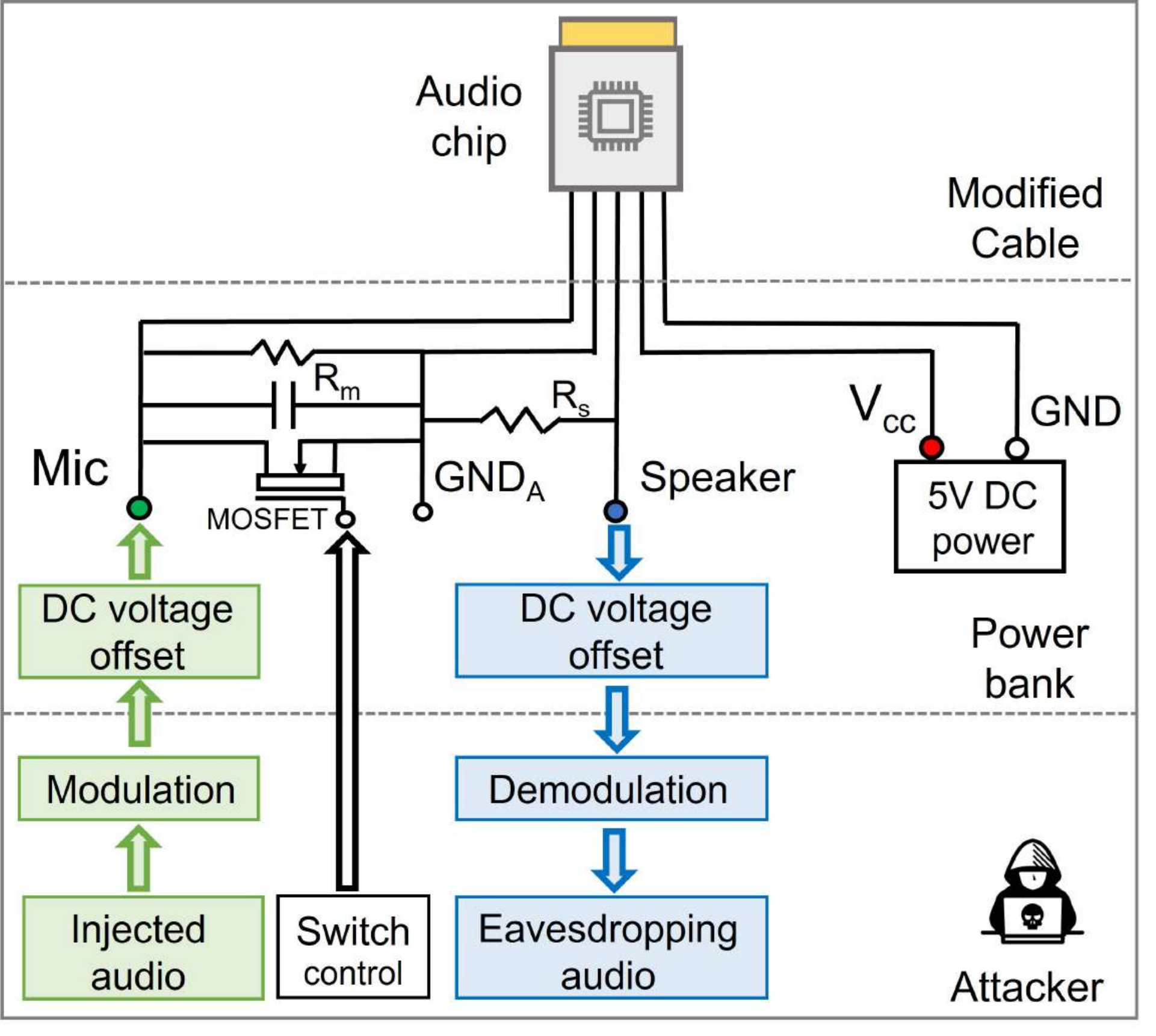}
    \caption{The system design of \ours.}
    \label{fig:system1}
    \vspace{-5pt}
\end{figure}

Fig.~\ref{fig:system1} illustrates the system design of \ours. 
After replacing the standard cables with specially designed cables described in Section~\ref{fakecable}, the attacker is able to manipulate the voltage on the microphone wire and monitor the voltage on the speaker wire.
The DC power in the power bank can charge the phone, and at the same time supply power for the hardware of \ours system. 
We add two resistors $R_{m}$ and $R_{s}$ between the microphone, speaker, and GND$_{A}$ wires to emulate the existence of a headphone. 
The resistance of $R_{m}$ and $R_{s}$ are 2,000 and 20 ohms, respectively.

\begin{figure*}
    \centering
    \subfigure[The original audio spectrogram.]{\includegraphics[height=3.4cm]{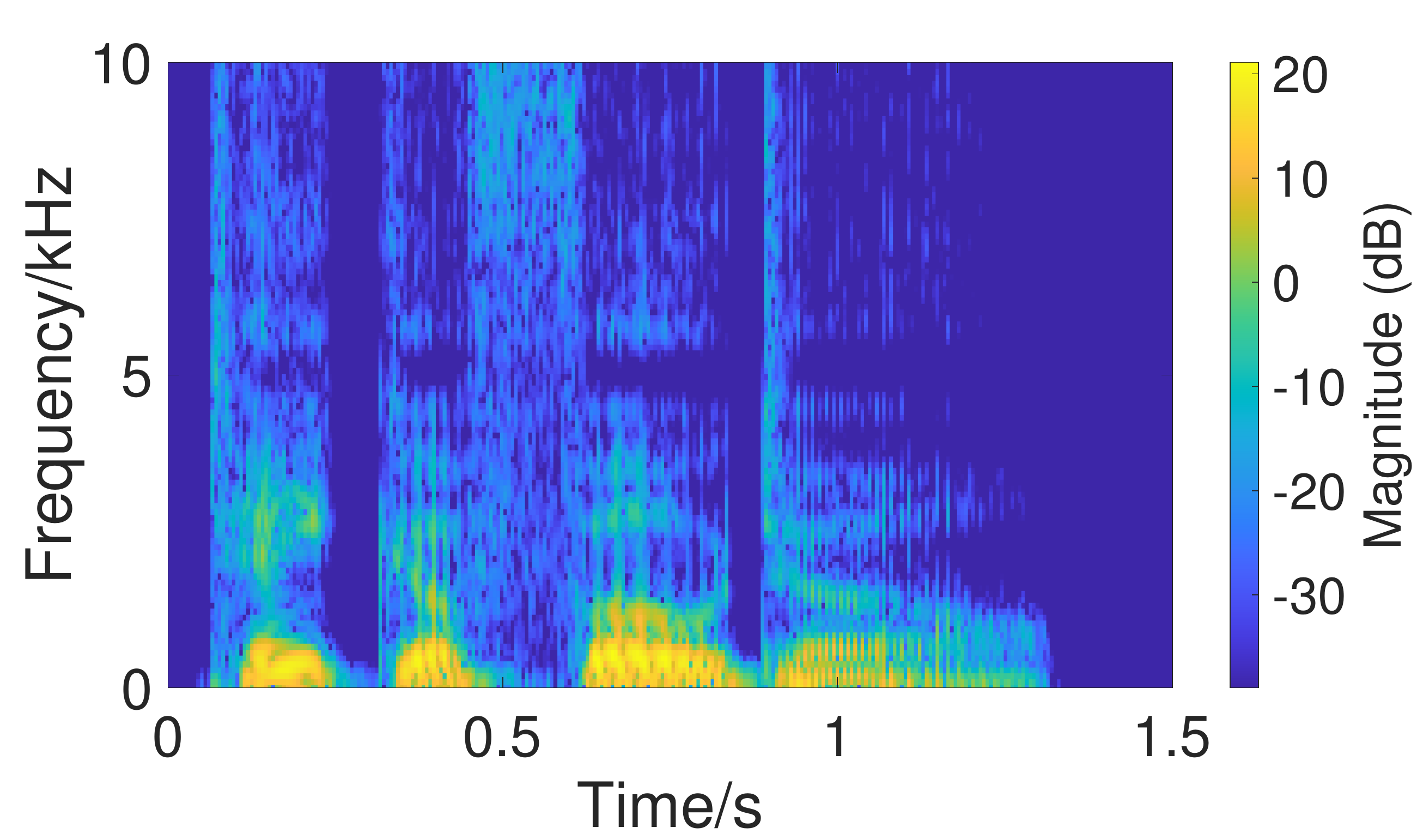}\label{fig:who clear}}
    \subfigure[The injected audio spectrogram (without using capacitor).]{\includegraphics[height=3.4cm]{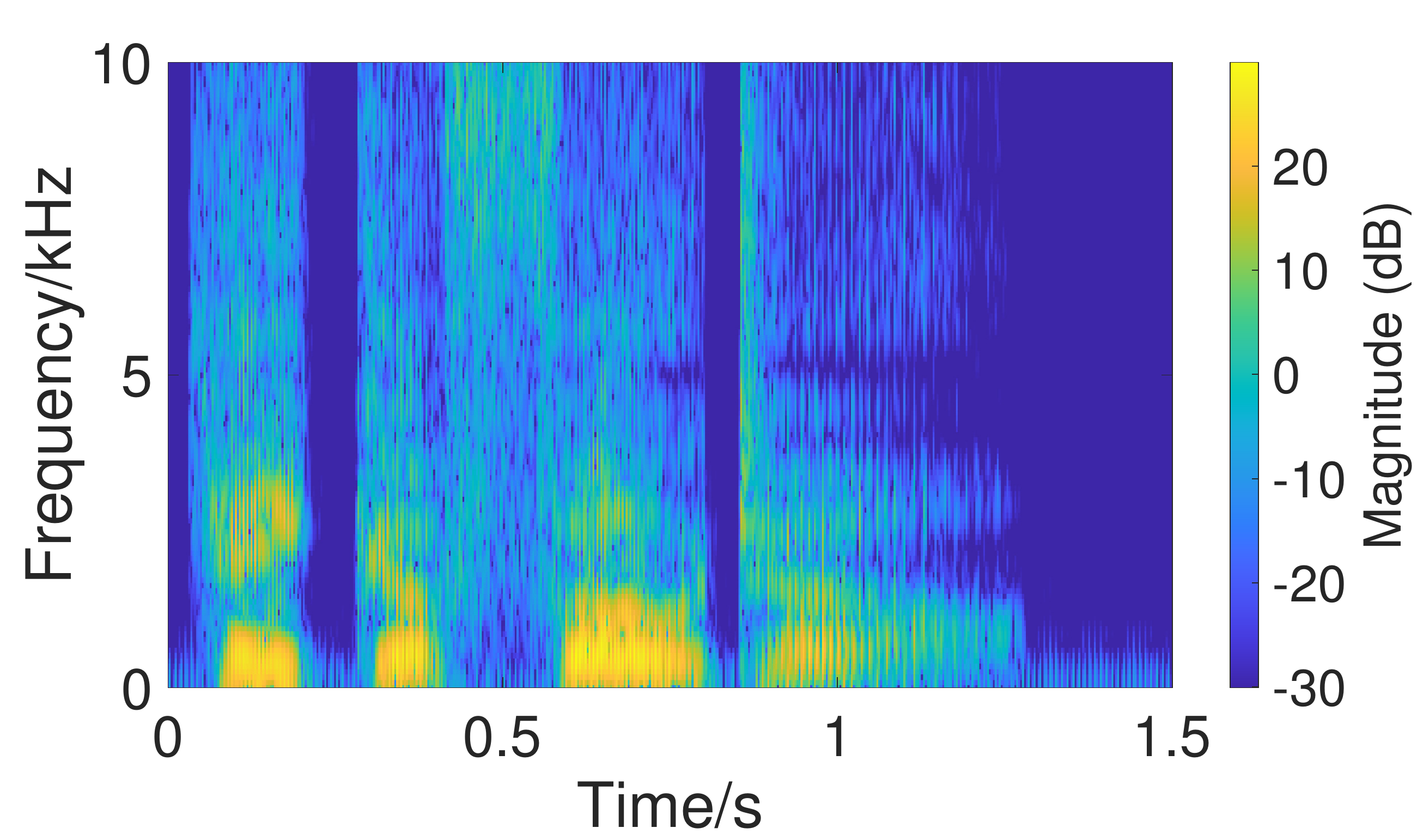}\label{fig: who noisy}}
    \subfigure[The injected audio spectrogram (with an additional capacitor).]{\includegraphics[height=3.4cm]{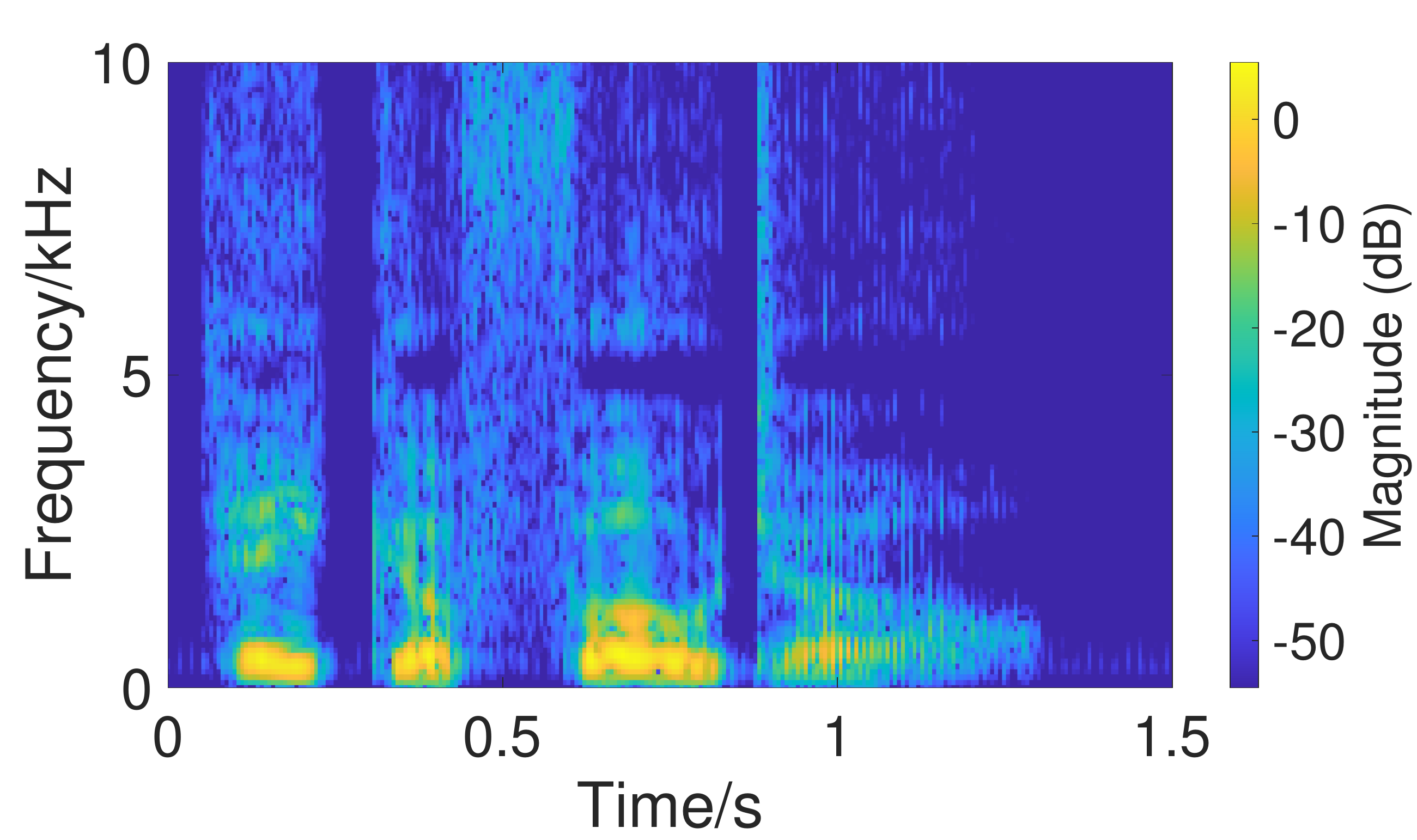}\label{fig: who capacitor}}
    \caption{The capacitor in \ours attack system can effectively reduce the noise in the injected audio.}
    \label{fig:adding capacitor}
    \vspace{-5pt}
\end{figure*}
\subsubsection{Voice Assistant Activation}
The first challenge of the \ours is to activate voice assistants without authorized user's voice.
Previous inaudible voice command injection attacks~\cite{yan2020surfingattack, zhang2017dolphinattack} require a collection of voice samples from the authorized user to generate specific commands such as ``Hey Siri" or ``Hello Google" to wake voice assistants.
However, if there is a lack of access to available authorized voice samples, such attacks become infeasible.

To address this challenge, we simulate the headphone press button function by manipulating the voltage in the charging cable. Our idea comes from the following observation: 
when the user is using a wired headphone, the voice assistants can be activated by pressing the button even with a locked smartphone. 
Therefore, the attackers can leverage the \emph{button pressing backdoor} to activate the voice assistant.
To replicate the press button operation, we add a MOSFET between the microphone and GND$_{A}$ wires.
Before injecting the malicious voice commands, the attacker activates the MOSFET to short the microphone and GND$_{A}$.
This operation will let the phone mistakenly believe that the user is pressing the button, leading to the activation 
the voice assistant.
Compared with other approaches on stealthy voice assistant activation, \ours has two advantages: 
first, 
the \ours attacker activates the voice assistant by electric signals, which are more resilient in noisy environments compared with voice command injection attacks;
second, \ours could bypass the speaker recognition system. 
After activating the voice assistant, the smartphone generally does not  verify the speaker's voice again. Therefore, the attacker can command the voice assistant using any voice.

\subsubsection{Inaudible Audio Injection}\label{subsec:inaudible}
After activation, the attackers can inject inaudible voice commands to the smartphone.
Eq.~(\ref{injection modu}) illustrates the injected signal modulation process, where $x_{i}(t)$ is the injected audio signal and $k$ is a factor to adjust the voltage range. 
Consider that the microphone capacitor has an initial voltage, we use an amplifier to add a DC offset $\Delta V_{in}$ ($\sim1.5$V) on the injected signal to compensate for the initial voltage of the microphone. The modulated voltage signal $V_{i}$ can be written as:
\begin{equation}
    V_{i}(t)=kx_{i}(t)+\Delta V_{in}. 
    \label{injection modu}
\end{equation}

However, 
the direct injection of the modulated voltage will generate a noisy injected audio.
Fig.~\ref{fig: who noisy} shows the spectrogram of an injected voice command ``take a photo". 
Compared with the original audio spectrogram in Fig.~\ref{fig:who clear}, the injected audio has substantial background noise. 
Such noise could degrade the audio quality and allow the listeners to identify the injected audio.

Fortunately, in our experiments, we observe that the headphone microphone's capacitor is not only used for generating the changing current, but it also functions as a signal smoother that smooths discrete voltage signals.
Therefore, 
we add an additional capacitor with similar capacitance 
to suppress the noise. 
Fig.~\ref{fig: who capacitor} shows the injected audio spectrogram after adding the capacitor, which is almost indistinguishable with the original audio spectrogram.  

\subsubsection{Inaudible Audio Eavesdropping}
When charged by the modified cable, the victim smartphone will play audio through a non-existent ``headphone" rather than a loudspeaker. 
Therefore, the attackers are able to capture the audio signals without accessing the audible sound.
Specifically, the attacker eavesdrops the audio signal by measuring the voltage of the speaker wire, as represented by blue boxes in Fig.~\ref{fig:system1}. Note that the modified cable has two wires for left and right speakers respectively, and \ours only needs to measure the voltage from one of them. Since the speaker wire carries analog signals, the attacker uses ADC to process and normalize the signals.  
Also, we use an amplifier to add an initial voltage offset $\Delta V_{out}$ ($\sim 1.5$V) to obtain an absolutely positive voltage $V_{o}(t)$ for the audio input, since the attacker's ADC can only process the signals with positive voltage.
Then, we can demodulate the audio signal $x_{e}(t)$ as follows: 
\begin{equation}
    x_{e}(t)=\frac{V_{o}(t)-\Delta V_{out}}{k},
    \label{eavesmodu}
\end{equation}
where $k=\textnormal{max}\{|V_{out}-\Delta V_{out}|\}$.
\subsection{System Design of \ourss}
\begin{figure}
    \centering
    \includegraphics[width=2.5in]{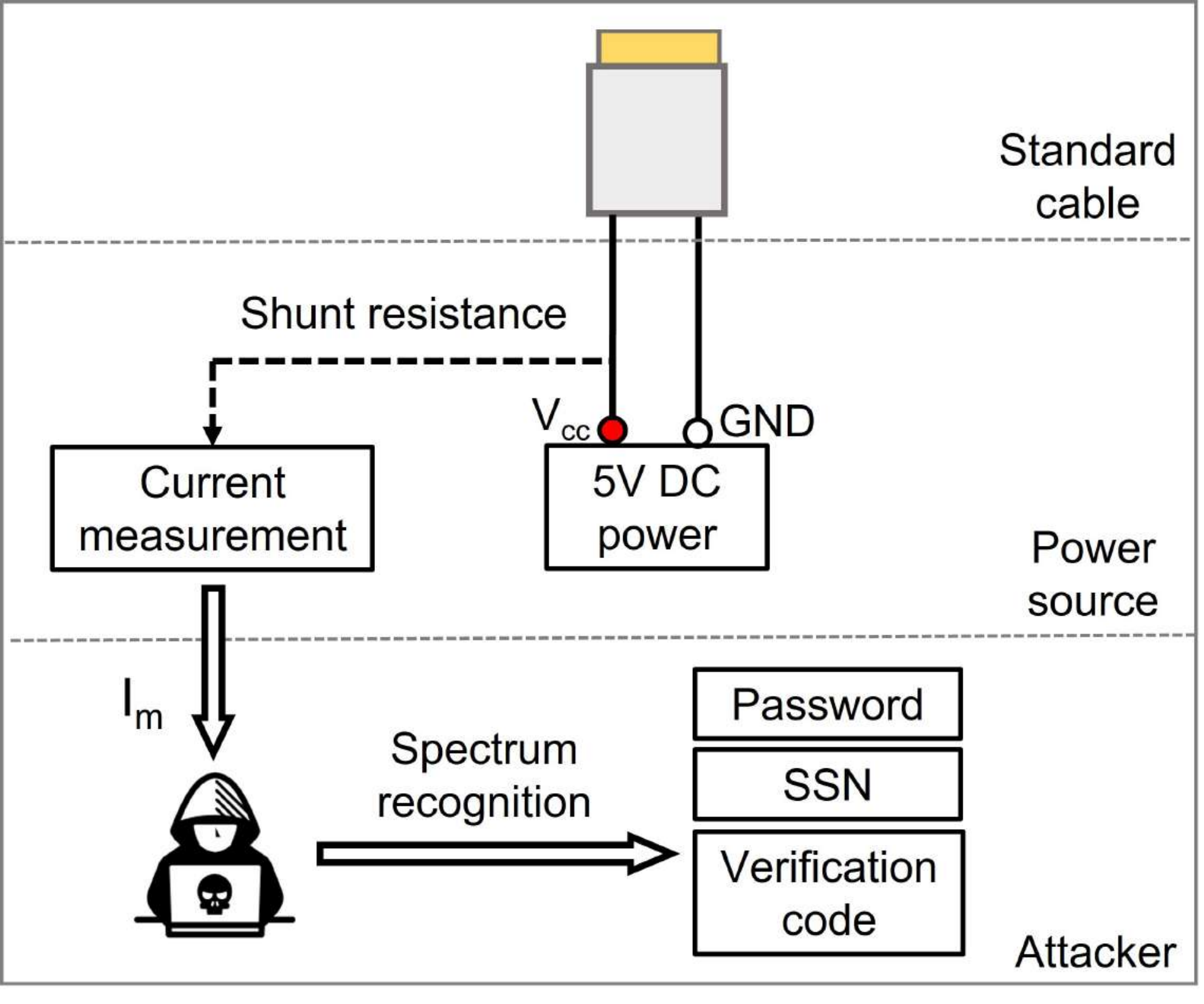}
    \caption{The system design of \ourss.}
    \label{fig:system2}
\vspace{-5pt}
\end{figure}

In this case, the victims charge their phones using their own standard cables. 
By passively monitoring the charging current, the attackers can eavesdrop private information through the power line side-channel.
Fig.~\ref{fig:system2} illustrates the system design of \ourss. 
Compared with \ours, \ourss system only needs to measure the charging current in the standard cable.
However, the demodulated audio only has a limited frequency band, which is distorted by strong background noise, making it incomprehensible for human ears. 
To resolve this challenge, we design a signal processing mechanism and apply a deep learning model to facilitate the recognition of the private information in the speech audio.
\subsubsection{Signal Processing}
After collecting the current measurement result $I_{m}(t)$, we use a high-pass filter to remove DC offset and low-frequency noise in the current signal, and recover a primitive audio $x_{n}(t)$.
%
\rev{Following the technique in~\cite{sami2020spying}, we apply spectral-subtraction to enhance the speech audio signal in $x_{n}(t)$.  
First, we obtain $X_{n}(\omega)$, the frequency domain spectra of $x_{n}(t)$ by Fast Fourier Transformation (FFT).
Meanwhile, by monitoring the idling smartphone charging current, we can estimate the signal strength of noise signal $N(\omega)$.
Then, we denoise $X_{n}(\omega)$ by:
$X_{c}(\omega)=X_{n}(\omega)-N(\omega)$, 
and transform the denoised frequency spectra $X_{c}(\omega)$ back to the time domain signal $x_{c}(t)$.}
\subsubsection{Digit Classification}
\begin{figure}
    \centering
    \includegraphics[width=8.5cm]{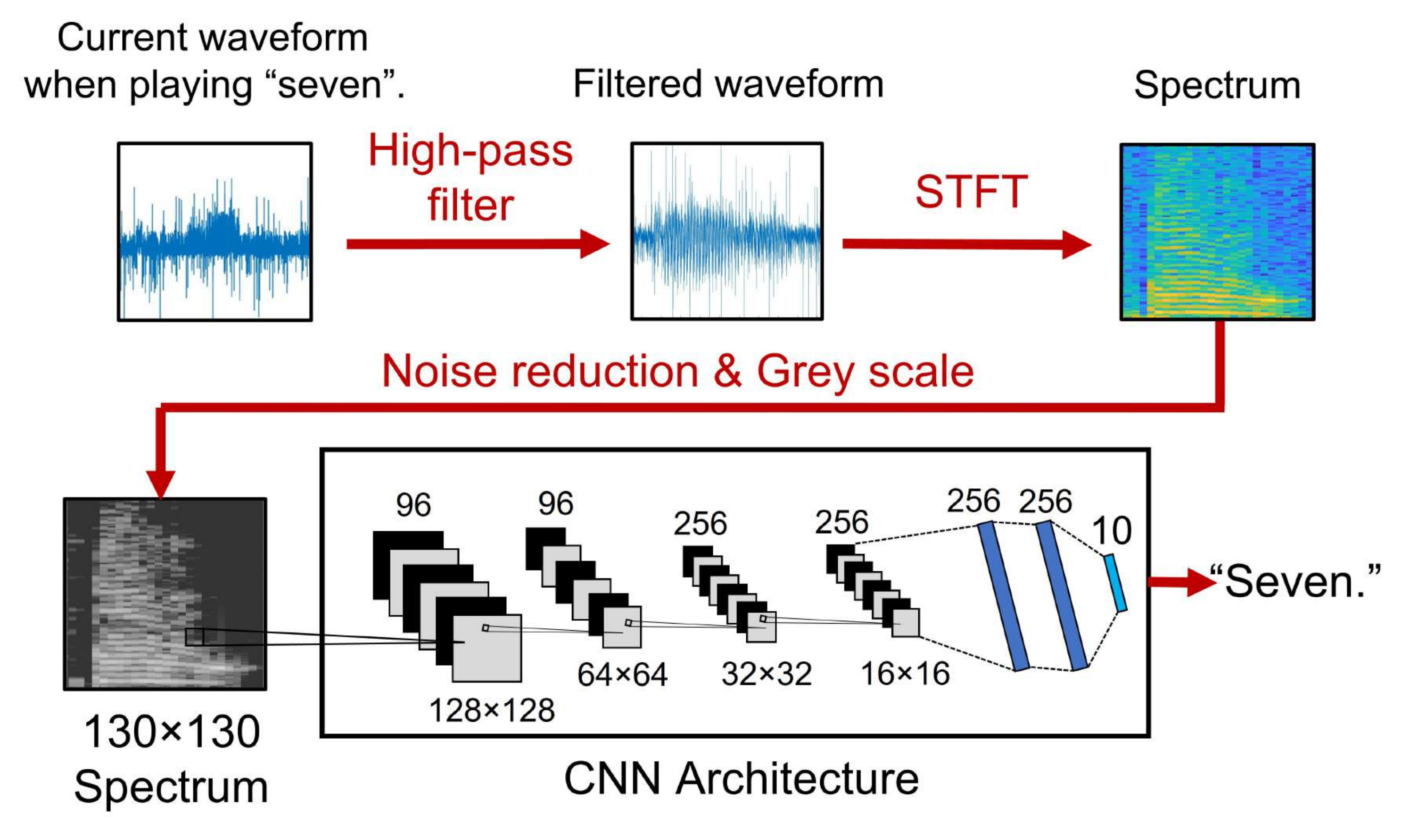}
    \caption{The CNN architecture of \ourss, which includes an input layer, two convolutional layers, two max-pooling layers, two dense layers, and an output layer.}
    \label{fig:cnn}
\vspace{-5pt}
\end{figure}
Unfortunately, after removing the background noise, the recovered audio $x_{c}(t)$ is still unrecognizable for either human or AI models. This is because only low frequency components in the audio (lower than 2 kHz) have been recovered from the current signals due to the signal loss. 
Our intuition is that the deep learning models such as CNNs can extract more convoluted patterns from voice signals, which can help recognize the speech using the low-frequency audio. 
Similar to the existing attacks~\cite{ba2020learning}, \ourss aims at realizing the digit recognition to extract sensitive information such as passwords, SSN numbers, and verification codes.

Fig.~\ref{fig:cnn} shows the CNN architecture of \ourss, which is used for classifying spoken digits from ``zero" to ``nine". 
\rev{The input to the CNN is a $130\times130$ spectrogram matrix of denoised audio signals generated by Short-Time Fourier Transform (STFT). 
The CNN model consists of two convolutional layers with ReLU activation and two $2\times2$ max-pooling layers.
Two dense layers with a dropout rate of 0.5 are used for improving the classification performance and preventing the overfitting~\cite{hertel2016comparing}.
Finally, a softmax layer outputs the probability distribution of ten digits.}
Using the trained model, the attacker can infer the spoken digits from the leaked speech audio. 
Compared with other speech recognition methods, the CNN architecture coherently learns from the time-domain and frequency-domain signals, which achieves high classification accuracy as shown in Section~\ref{evaluation}.
\section{Evaluation}\label{evaluation}
\subsection{\ours Attack Evaluation}
\subsubsection{Experiment Setup}
\begin{figure}
    \centering
    \includegraphics[width=8.5cm]{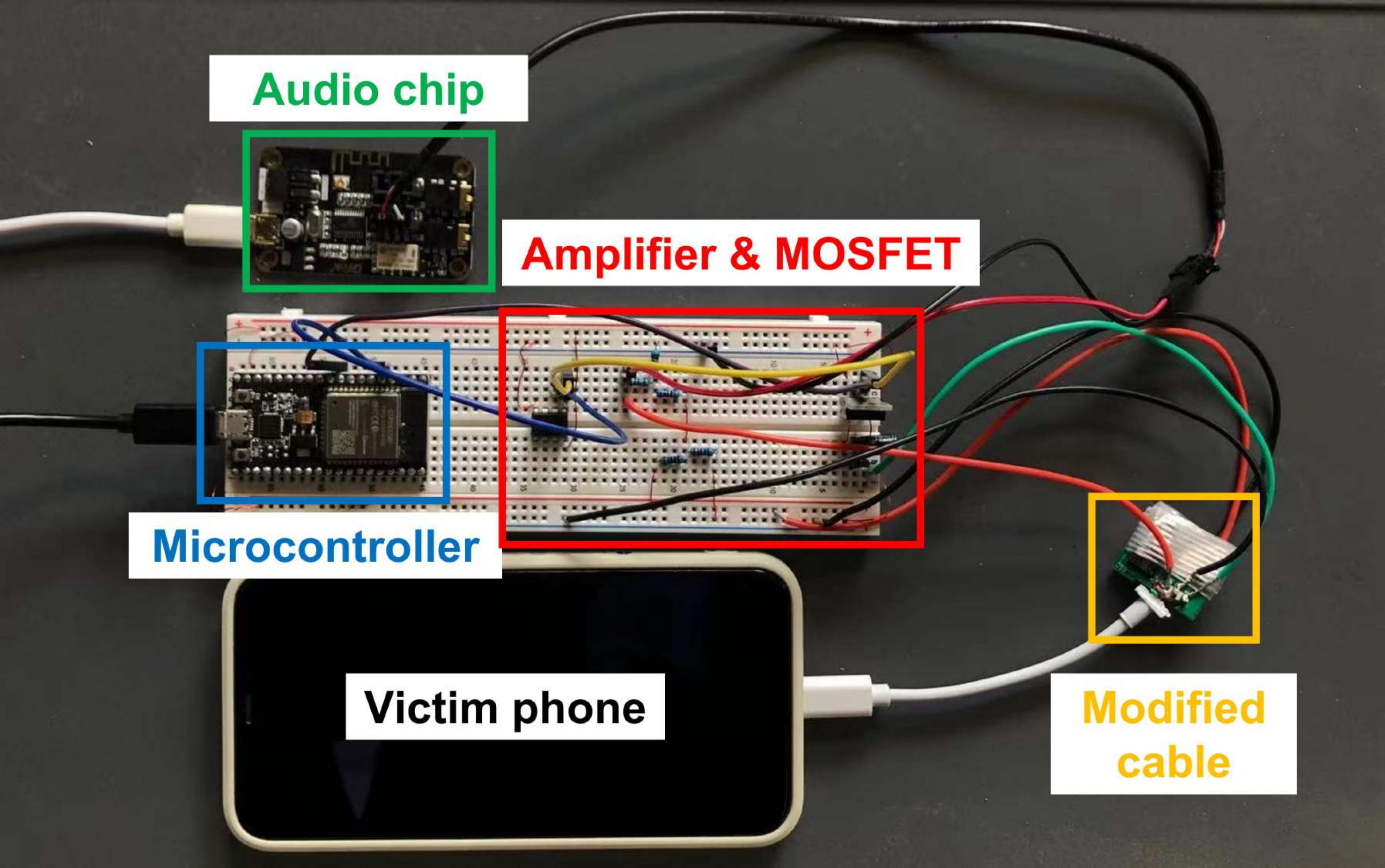}
    \caption{Low-cost and portable experiment setup of \ours attack. The hardware devices are small enough to be hidden in a power bank, and the modified cable has the same outlook as a standard cable.}
    \label{fig:setup}
\end{figure}

\begin{table*}

    \centering
    \caption{\textnormal{Experiment devices, operating systems, and microphone sampling frequencies. We test three components of \ours attacks including voice assistant activation (Act.), inaudible voice command injection (Inj.), inaudible audio eavesdropping (Eav.). $f_{s}$: the sampling frequency of the smartphone microphone; SNR: Signal-to-Noise ratio of injected audio; ASR: injection attack success rate of \ours.}}
    \begin{tabular}{ccccccccccc}
       \toprule
    \multirow{2}*{Num.}&\multirow{2}*{Manufacturer}&\multirow{2}*{Model} & \multirow{2}*{OS/Ver.} & \multirow{2}*{Assistants} &\multirow{2}*{\tabincell{c}{$f_{s}$\\(kHz)}}& \multicolumn{3}{c}{\ours}&\multirow{2}*{\tabincell{c}{SNR\\(dB)}}&\multirow{2}*{\tabincell{c}{ASR}}\\
   \cline{7-9}
    ~&~&~&~&~&~& Act.&Inj. & Eav. &~&~\\ 
    
    \midrule
    1&Apple& iPhone 5s & iOS 12.5 &Siri&44.1&\checkmark&\checkmark&\checkmark & 19.7 & 100\%\\
    \midrule
    2&Apple& iPhone X  & iOS 14.5 & Siri &48.0&\checkmark&\checkmark&\checkmark& 21.3 & 100\%\\
    \midrule
    3&Huawei& Honor 10 & Android 9.0 & Google &48.0 &\checkmark&\checkmark&\checkmark & 20.4 & 100\%\\
    \midrule
    4&Xiaomi& MI 8 Lite & Android 9.0 & Google  &44.1 &\checkmark&\checkmark&\checkmark & 18.9 & 100\%\\
    \midrule
   5& Xiaomi& Pocophone & Android 9.0 & Google  &48.0 &\checkmark&\checkmark&\checkmark & 21.8 & 100\%\\
    \midrule
   6& Samsung& Note 10 & Android 10.0 & Google  &44.1 &\checkmark&\checkmark&\checkmark & 21.2 & 100\%\\
    \midrule
   7& Samsung& Galaxy S9 & Android 10.0 & Google &44.1 &\checkmark&\checkmark&\checkmark & 20.1 & 100\%\\
    \midrule
    8&Google& Pixel 1 & Android 10.0 & Google & 44.1 &\checkmark&\checkmark&\checkmark & 19.3 & 100\%\\
    \midrule
   9& Google& Pixel 4XL & Android 11.0 & Google  &32.0 &\checkmark&\checkmark&\checkmark & 15.4 & 100\%\\
    \bottomrule
    \end{tabular}
    \label{table:devices}
\end{table*}
In the experiments, we evaluate the \ours attack on 9 different smartphones from 5 mainstream manufacturers including Apple, Google, Samsung, Huawei, and Xiaomi. 
The experiment setup is shown in Fig.~\ref{fig:setup}.
An ESP-32 board with WiFi and Bluetooth modules is used to control the MOSFET and measure the voltage from the speaker wire. A Bluetooth audio chip injects the modulated audio commands to the victim phone, and 
an LM-358 dual-channel amplifier is used to apply the DC voltage offset.

\subsubsection{\ours Injection Performance}\label{inject_quality}
To evaluate the performance of \ours injection attack, we use Google WaveNet API~\cite{oord2016wavenet} to generate 20 voice commands, and each command contains $3\sim8$ words.
After activating the voice assistant, we inject the voice commands to each victim smartphone, and repeat the experiment for 10 times.
Then, we calculate the signal-to-noise ratio (SNR) of the voice command recordings.

The results are listed in Table~\ref{table:devices}. 
For all the victim smartphones, the average injection audio SNR is higher than $15$ dB, which can be clearly perceived by human ears~\cite{stuart1994noise}.
Also, we notice that the injected audio SNR is related to the microphone sampling frequency $f_{s}$ of the victim smartphone.
For example, for Samsung Note 10 ($f_{s}$ = 44.1 kHz), the injected recordings have higher average SNR value than Pixel 4XL ($f_{s}$ = 32.0 kHz). 
The attacker can further improve the injected audio volume by adjusting the amplification factor $k$ in Eq.~(\ref{injection modu}).
It is worth noting that a larger $k$ may degrade the performance of \ours injection if the current in the microphone is beyond the smartphone sampling range.
In our experiment, we set $k$ = 0.1 to balance the audio quality and SNR value.



Next, we repeat the experiment and test if these injected voice commands can be recognized by the voice assistants.
we list the attack success rate (ASR) result of \ours injection attack in the last column of Table~\ref{table:devices}. 
Surprisingly, in spite of different hardware design and sampling frequency, all victim smartphones are vulnerable to \ours injection attack.
For all victim smartphones, \ours injection attack can compromise their voice assistants with 100\% success rate, which outperforms all state-of-the-art inaudible command injection attacks.
\subsubsection{\ours Eavesdropping Performance}\label{eavesperfor}
To evaluate \ours eavesdropping attack, we play 100 human speech samples from TIMIT dataset~\cite{zue1990speech} with the maximum volume setting of the victim smartphone. 
Meanwhile, the ESP-32 board works as an ADC to measure the voltage output from the speaker cable with $10$ kHz sampling frequency.
For comparison, we play these speech samples with the loudest volume in a quiet environment (noise level $\leq$ 25 dB), and use an iPhone 8 to record the audio $30$ cm away from the victim smartphone.
Subsequently, we compare the eavesdropping performance of \ours against a normal recording.

\begin{figure}
    \centering
    \includegraphics[width=3in]{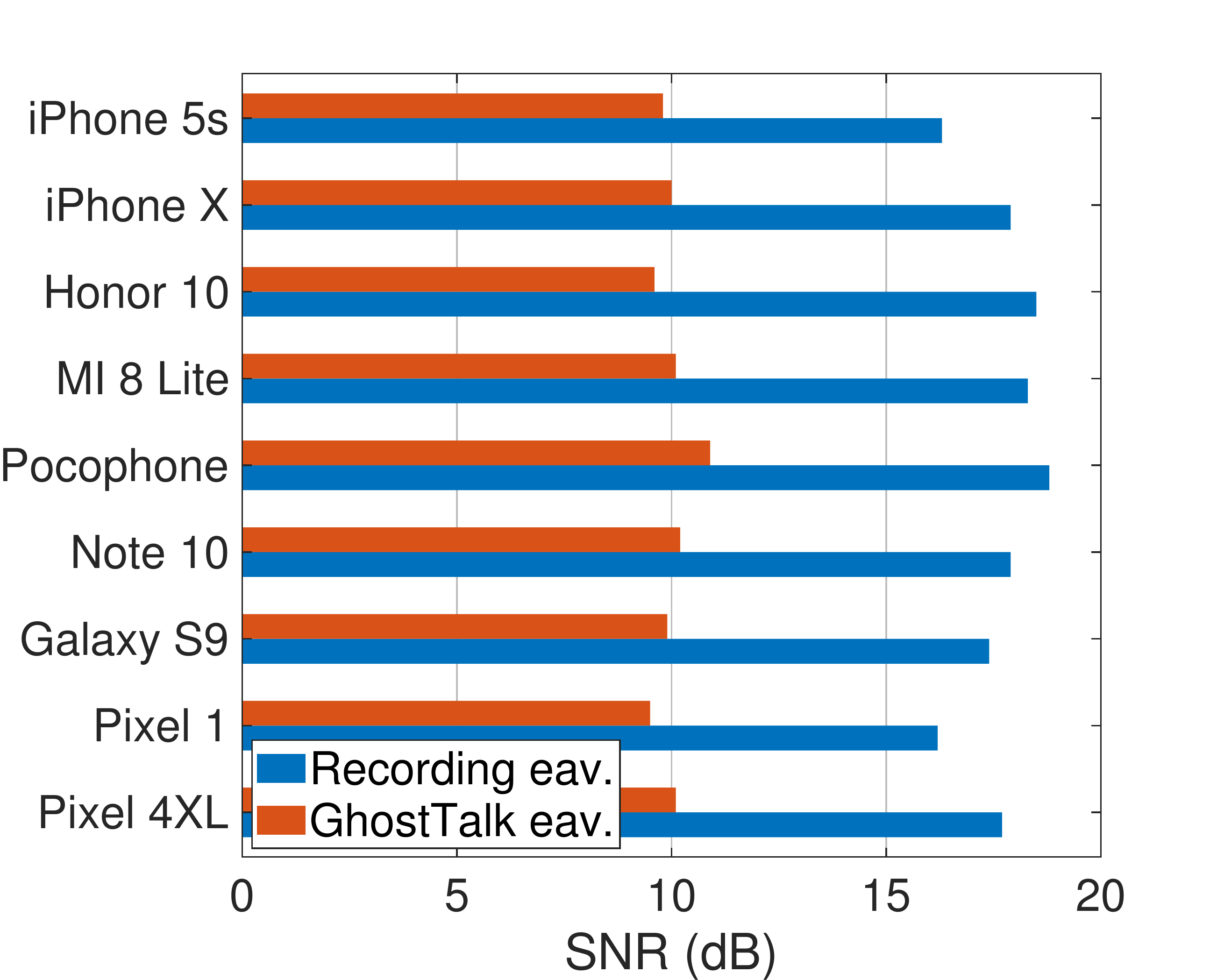}
    \caption{SNR comparison of normal recording and \ours eavesdropping.}
    \label{fig:eav_snr}
\end{figure}
 
For normal recording, the smartphone loudspeaker's output power determines the SNR of eavesdropped audios. For \ours attack, the recovered audio SNR is limited by the voltage range of the speaker wire and the sampling frequency of ADC.
Fig.~\ref{fig:eav_snr} shows the audio SNR comparison of recording eavesdropping and \ours eavesdropping.
For all the victim smartphones, \ours eavesdropping has lower average SNR values because of the low sampling frequency and restricted voltage amplitude.
However, since most of human voice spectrogram is below $5$ kHz~\cite{stevens2000acoustic}, our eavesdropping attack can still recover clear human speech audios.

To further evaluate the eavesdropping audio quality and clarity, we use Google Speech-to-Text API~\cite{googlestt} to recognize the speech contents in the eavesdropped audios.
For both recording eavesdropping and \ours eavesdropping, Google Speech-to-Text API can accurately recognize approximately 95\% of words in the speech audios regardless of the SNR.
The result demonstrates that \ours could eavesdrop the audios through the power line, and attain the same audio clarity as an eavesdropping attack with normal audio recording in a quiet environment.
\subsubsection{Human Study}\label{humanstudy}
As shown in Section~\ref{threatmodel}, the attackers can launch ghost calls using \ours attack, i.e., 
the attacker can initiate a phone call by injecting voice commands, and ``speak" with the victim's voice.
To deceive the human ears, the injected audios by \ours are supposed to have the same quality as natural human speech.
Therefore, to verify the feasibility of ghost call attack, we design a human study experiment to test if human ears can distinguish natural and injected human speech audios.

First, we use an iPhone X to record ten natural human speech samples from ten different speakers.
Then, we launch the \ours attack to inject these audio samples to the same smartphone, and obtain 10 corresponding injected human speech samples with the same speech contents. 
\rev{
We then normalize all the benign and injected voice samples to eliminate the amplitude or length disparity.

In total, 20 volunteers (12 males and 8 females) participate in our human study.
As a baseline, the volunteers are first requested to listen to two sets of voice samples as training examples. 
In each set, one sample is natural human speech, and the other one is an injected human speech sample from the circuit without the capacitor (see Fig.~\ref{fig: who noisy}).
Since the injected samples present audible noise and high-frequency distortion, all the listeners can correctly recognize the injected samples. We also verbally explain to the volunteers that the injected samples may contain additional noise, and may be subject to frequency distortion in comparison with the natural audio samples.

Next, for each question set, we have one natural human speech sample and one injected sample from the \ours injection  (sample A \& B).
After listening to one set, the volunteers need to select whether the samples can be distinguishable or not. Then, the volunteers will select the likely injected sample.

In the end, we collected $200$ answers. Table~\ref{tab: human study} summarizes the human study results, among which 150 answers depict the two samples as ``indistinguishable", and the rest 50 answers claim the opposite. Out of the 50 answers, only 30 of them successfully pinpoint the injected sample. 

Along with the sample recognition study, we also conduct a survey of the volunteers. 
First, the volunteers are requested to mark their answers if they randomly guess the answers, and if not, they are asked to explain their selections. 
Out of the 50 ``distinguishable" answers, 32 are derived by random guessing.
Moreover, more than 70\% of volunteers who provide deterministic answers claim that there is subtle audible noise in the injected samples, which is likely introduced by the attack circuit. 
Table~\ref{tab: human study} presents the accuracy of both random guess and deterministic answers. 
In fact, the deterministic answers, despite the responders’ confidence, achieve very similar accuracy as the random guessing. 
Meanwhile, 17 out of 20 volunteers indicate that they will be incapable of recognizing the injected voice during a real phone call. 

}
\begin{table}
\small
\centering
\begin{tabular}{cccc}
\toprule
                                 & Type of answer & Number & Accuracy                 \\
\midrule
\multirow{4}{*}{Distinguishable} & Random guess \checkmark     & 19     & \multirow{2}{*}{59.4\%}  \\
                                 & Random guess \ding{53}       & 13     &                          \\
\cmidrule{2-4}
                                 & Deterministic \checkmark  & 11     & \multirow{2}{*}{61.1\%}  \\
                                 & Deterministic \ding{53}  & 7      & \\
\midrule

Indistinguishable                & N/A            & 150    & N/A                      \\
\midrule
Overall & N/A & 200 & 15\%\\
\bottomrule
\caption{\ours human study results. \checkmark and \ding{53} respectively indicate correct and wrong answers. Only 15\% of answers correctly select the injected voice sample, and most samples are indistinguishable for the volunteers. In addition, the deterministic answers have a similar accuracy as the randomly guessed answers.}
\label{tab: human study}
\end{tabular}
\vspace{-5pt}
\end{table}


\subsubsection{Liveness Detection Robustness}
To defend against replay attack and inaudible voice command injection, the voice assistants can apply liveness detection models to recognize the maliciously injected voice commands.
To evaluate the attack robustness of \ours against liveness detection models, we inject 100 human speech samples from TIMIT dataset to an iPhone X, and input the injected recordings to three liveness detection models.
The first model is ASVSpoof~\cite{kinnunen2017asvspoof}, the baseline liveness detection model of ASVSpoof 2017 challenge. ASVSpoof mainly considers the constant-Q cepstral coefficients (CQCC) features in the voice and leverages Gaussian Mixture Model (GMM) to separate the natural and replayed human speech. 
The second model, STC~\cite{lavrentyeva2017audio}, is the best model in ASVSpoof 2017 challenge, which incorporates a Light Convolutional Neural Network (LCNN) to detect replay attacks.
The third model, Void~\cite{ahmed2020void}, is a state-of-the-art liveness detection system that detects replayed samples and inaudible voice commands using spectrogram delay patterns, peak patterns, and Linear Prediction Cepstrum coefficient (LPCC) features.

We use the reported results for replay and inaudible voice attacks in the respective defense studies, and evaluate the robustness of \ours against the liveness detection models.
The evaluation results are presented in Table~\ref{table:liveness}.
For the replay attack using a loudspeaker, only a few samples can bypass the liveness detection systems. 
Moreover, Void could accurately recognize all audio samples from inaudible voice command injection. 
As for \ours injection attack, when we inject audio samples with 48 kHz sampling frequency, all of the samples can bypass ASVSpoof and STC models, which is understandable since \ours injection occurs from the power line rather than the loudspeaker.
It is worth noting that only 40\% of the injected samples could successfully fool the Void system, which is likely due to the low-frequency patterns of the injected samples captured by the Void model. 
In response, we decrease the injected audio sampling rate from 48 kHz to 16 kHz, which effectively distorts the low-frequency patterns of the injected audio. 
As a result, 81\% of injected samples can pass the Void model. 
On the other hand, the downsampling process also distorts or discards some high-frequency components, which degrades the error rate of the STC model to 63.0\%. 
In summary, by tuning the injected audio sampling frequency, \ours injection attack can successfully bypass different liveness detection models.

\begin{table}
\small
    \
    \caption{\textnormal{The error rate of liveness detection systems against replay attack, inaudible voice command injection, and \ours injection attack.}}
    \begin{tabular}{cccc}
       \toprule
    Attacks & ASVSpoof~\cite{kinnunen2017asvspoof} & STC~\cite{lavrentyeva2017audio} & Void~\cite{ahmed2020void}\\
    \midrule
    Replay attack & 24.77\% & 6.73\% & 8.7\%\\
    Inaudible attack & N/A & N/A & 0\%\\
    \textbf{\ours (48kHz)} & \textbf{100\%}  & \textbf{100\%} &\textbf{40.0\%}\\
    \textbf{\ours (16kHz)} &\textbf{100\%} & \textbf{63.0\%} &\textbf{81.0\%} \\
    \bottomrule
    \end{tabular}
    \label{table:liveness}
    \vspace{-5pt}
\end{table}


\subsection{\ourss Attack Evaluation}
\subsubsection{Experiment Setup}
We evaluate \ourss attack on the smartphones listed in Table~\ref{table:accuracy}.
The victim smartphones are charged by a $5$V/$1$A DC power source, and the charging cables are standard Lightning or USB-C cables. 
An ESP-32 board is used to measure the charging current fluctuation with $8$ kHz sampling frequency. 

\subsubsection{Data Collection} 
We train the CNN classifier with the Free Spoken Digit Dataset (FSDD)~\cite{jackson2018jakobovski} consisting of 3,000 utterances from ``zero" to ``nine". 
We also collect 300 utterances from 15 speakers (8 males and 7 females, and each of them speaks 10 digits twice) and extract the leaked speech audio as the test dataset.
Then, we classify the denoised voice samples by recognizing their spectrogram patterns under $2$ kHz.

\subsubsection{Digit Classification Performance.}\label{digitclass}
\begin{table}
\small
    \centering
    \caption{\textnormal{Smartphone hardware information, leaked audio signal SNR, and digit classification accuracy of \ourss.}}
    \begin{tabular}{ccccc}
       \toprule
    Model & \tabincell{c}{Charging\\port}&Loudspeaker &SNR (dB) & Accuracy\\
    \midrule
    iPhone 5s & Lightning& Single&5.41 & 93.0\%\\
    iPhone X & Lightning& Dual&4.75 & 92.7\%\\
    Honor 10 & USB-C&Single& 5.75 & 93.3\%\\
    MI 8 Lite& USB-C & Single&4.93 & 92.7\%\\
    Note 10& USB-C & Dual&4.46 & 91.0\%\\
    Galaxy S9& USB-C &Dual& 4.21 &  90.7\%\\
    Pixel 1& USB-C & Single&3.83 & 89.7\%\\
    Pixel 4XL& USB-C & Dual &3.72 & 90.0\%\\
    \textbf{Pocophone} & \textbf{USB-C}& \textbf{Dual}& \textbf{1.51} & \textbf{36.0\%}\\
    \bottomrule
    \end{tabular}
    \label{table:accuracy}
\end{table}
Table~\ref{table:accuracy} lists the average leaked audio SNR of different smartphones and the spoken digits classification accuracy.
The average SNR values vary substantially across different phone models due to the firmware and system difference. 

The results show that \ourss achieves satisfactory classification performance on 8 out of 9 victim smartphones.
For the smartphones with stronger leaked audio signals, such as Honor 10, iPhone 5s, and iPhone X, \ourss can achieve 92\% or higher classification accuracy, which is significantly higher than random guessing (10\%). 
For smartphones with weaker audio leakage, like Pixel 4XL, the accuracy descends. 
A special case is that \ourss fails to classify most of the spoken digits from Pocophone as most of leaked audio signals are overwhelmed by the ambient noise. 
\rev{This exception may be attributed to the weaker loudspeaker power.
Another possible explanation is that Pocophone's operating system (OS) or firmware runs with a higher power consumption that introduces excessive noise into the charging current.}
\begin{figure}
    \centering
    \includegraphics[width=8cm]{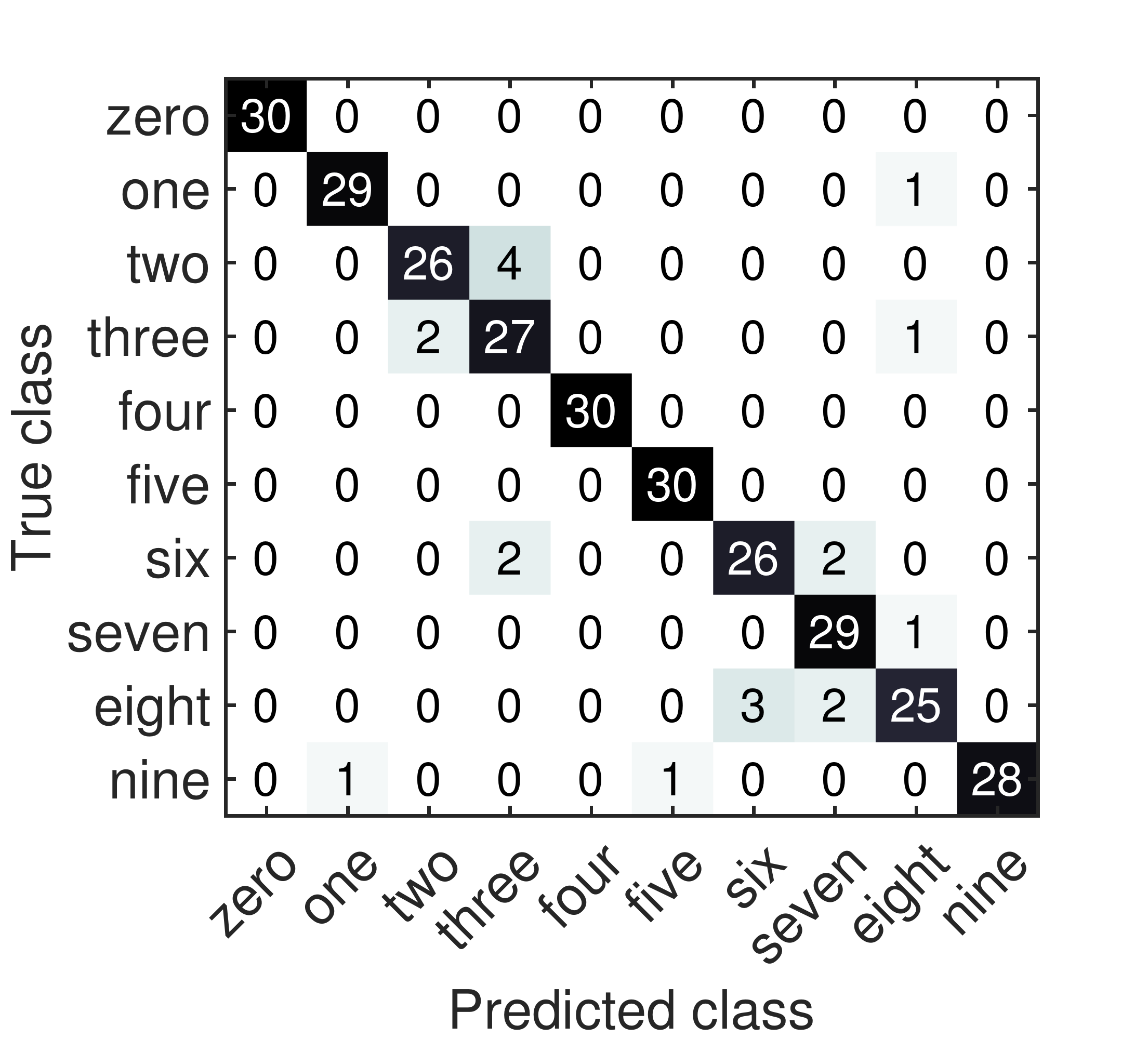}
    \caption{Spoken digit classification confusion matrix (with Honor 10 smartphone).}
    \label{fig:confusion_matrix}
\end{figure}

Fig.~\ref{fig:confusion_matrix} shows the digit classification confusion matrix (with counts of cases) from Honor 10 smartphone.
From the confusion matrix, we notice that 
there are false predicted samples between ``two" and ``three", and ``eight" is frequently misclassified as ``six" or ``seven", because these utterances have similar patterns in the low-frequency band. 
Due to the low SNR of leaked audio signal and frequency band limitation, the extracted digit utterances have lower distinguishability compared with the original utterances, which impacts the classification accuracy.

\subsubsection{\ourss Performance under Different Volume Settings}
\begin{figure}
    \centering
    \includegraphics[width=8cm]{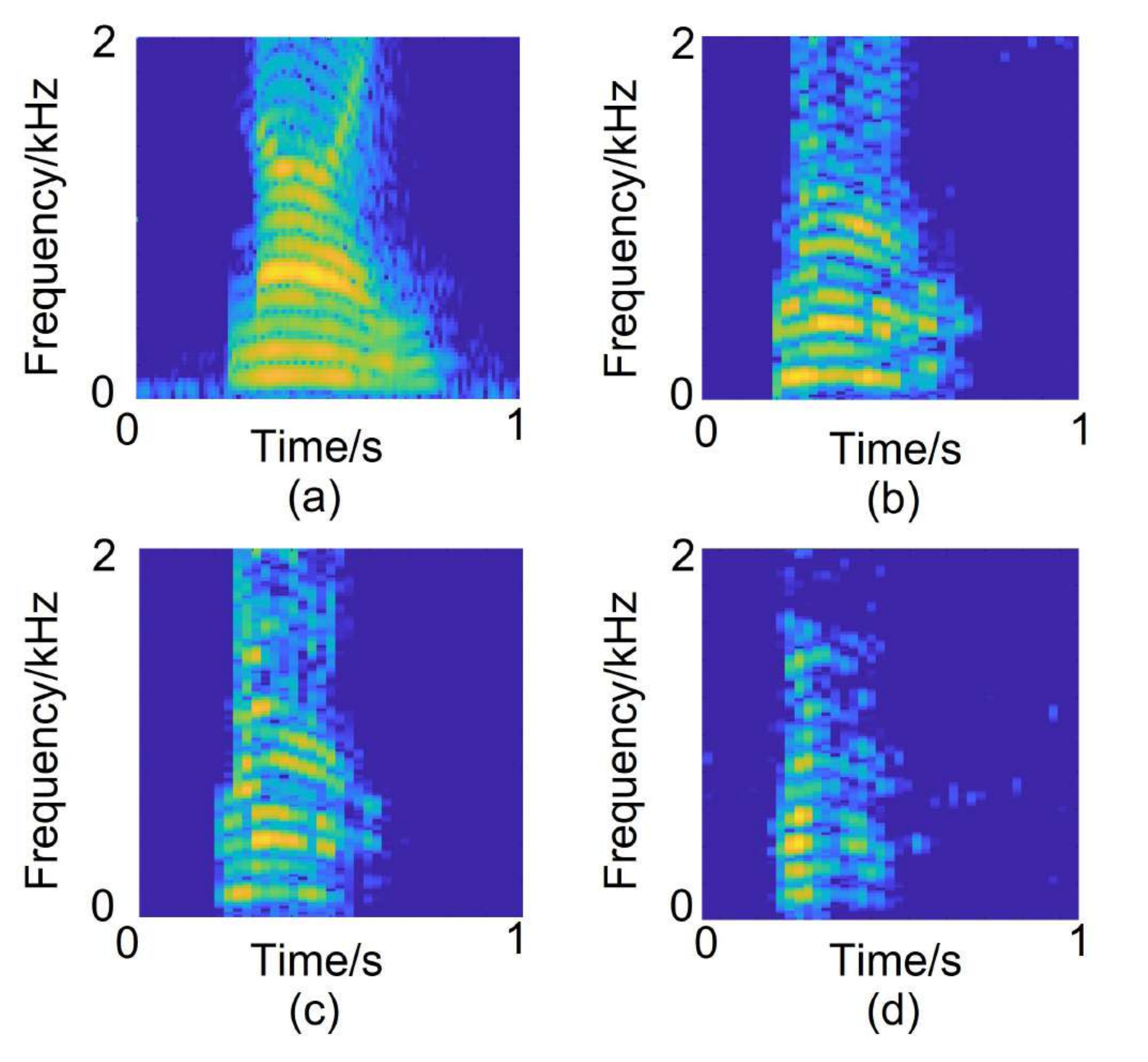}
    \caption{spectrogram comparison of original audio and leaked audios under different volume settings. (a) is the spectrogram of original utterance ``nine", and (b), (c) and (d) are leaked audio spectra when the volume level is 100\%, 75\% and 50\%, respectively.}
    \vspace{-5pt}
    \label{fig:spectrograms}
\end{figure}
\begin{figure}
    \centering
    \includegraphics[width=8cm]{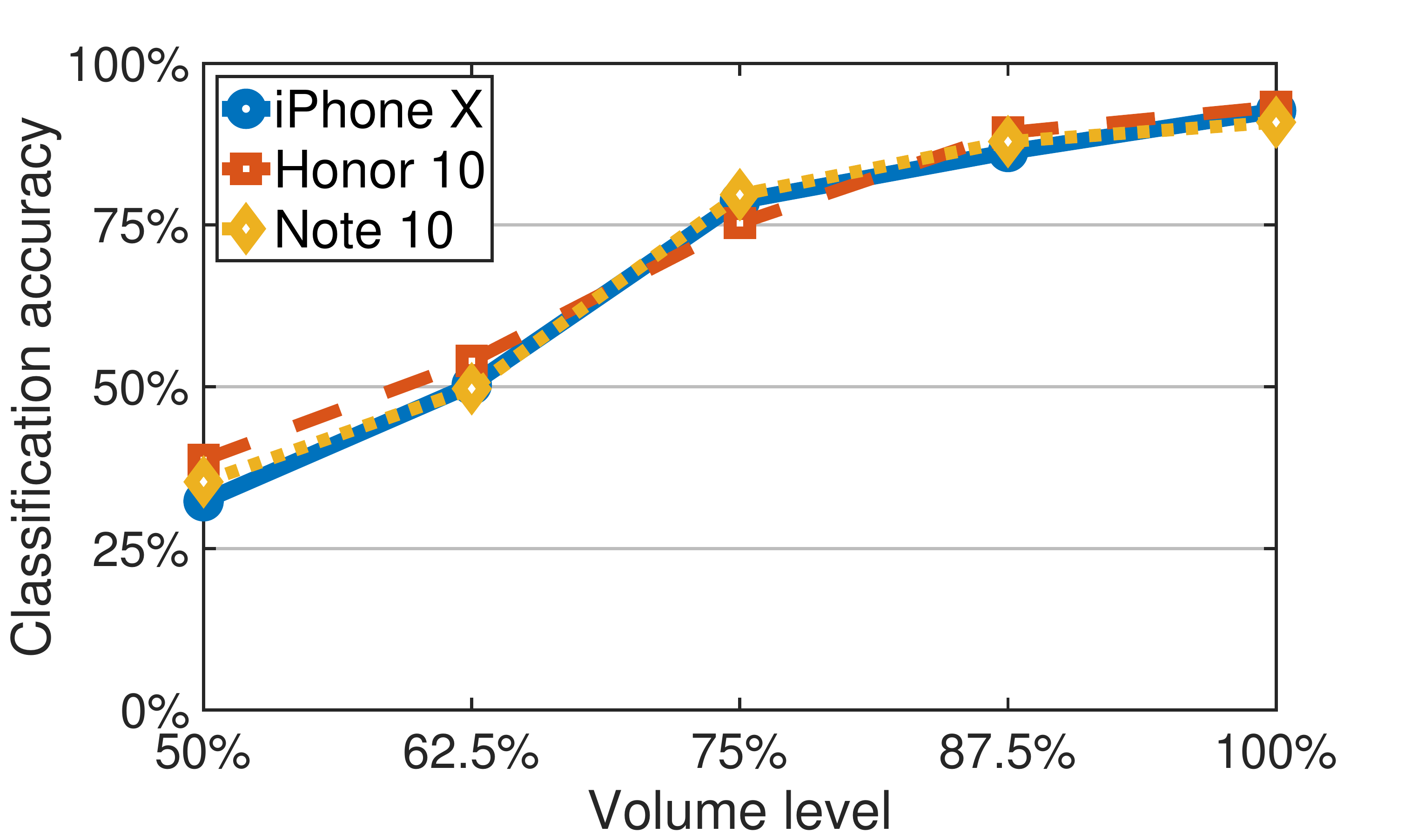}
    \caption{Classification accuracy under different volume settings.}
    \vspace{-5pt}
    \label{fig:volume}
\end{figure}

As we illustrate in Section~\ref{motivation}, the leaked audio signal in charging current is originated from the power side-channel.
When the user turns down the volume, the loudspeaker power consumption decreases which in turn leads to the drop of the SNR of leaked audio signals.
Fig.~\ref{fig:spectrograms} shows the original utterance and leaked audio spectra under different volume settings of Huawei Honor 10. 
When the audio is played with the maximum volume, most of spectrogram patterns can be well recovered after denoising. 
If the volume is reduced to 75\%, a portion of the patterns get lost or distorted after denoising.
With 50\% volume level, only the strongest frequency components remain in the spectrogram with most of patterns disappeared.

To evaluate the digit classification performance under different volume settings, we play the testing spoken digit audios on three victim smartphones including iPhone X, Honor 10, and Note 10.
Each smartphone has 16 volume levels.
We start with volume 100\% (level 16) and repeat the experiment after tuning the volume.

The digit classification accuracy under different volume settings is presented in Fig.~\ref{fig:volume}.
For all the three victim phones, the classification accuracy drops when  the volume is down.
When the volume is 75\% (level 12), the classification performance slightly degrades since most of utterances are still distinguishable.
However, when the volume is set as 50\% (level 8), the classification accuracy declines more drastically, i.e., 
only 35\% spoken digits can be correctly classified.
In a lower volume setting, the classification results approximate that of random guessing.

\subsection{Robustness Evaluation}
\subsubsection{\ours Injection Robustness}\label{injection_robustness}
Most of the existing voice attacks are susceptible to other acoustic interference, such as environmental noise, human conversation, and loud music. 
In the extreme, strong background noise will jam the microphone and block the voice command injection and audio eavesdropping. 

\begin{figure}
    \centering
    \includegraphics[width=8cm]{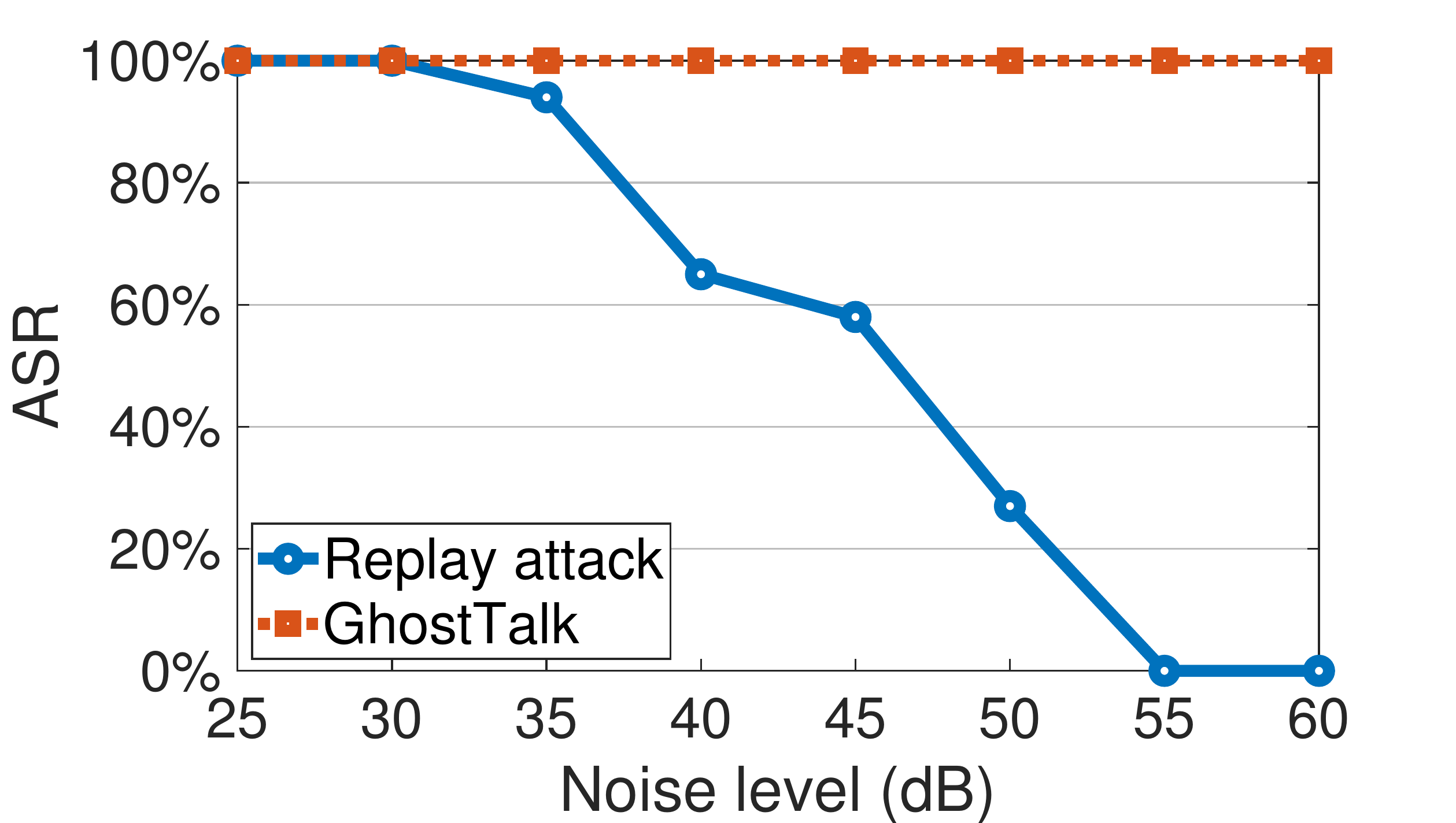}
    \caption{ASR comparison of replay attack and \ours injection attack in different noisy environments.}
    \label{fig:inj_rob}
\end{figure}

To verify \ours attack performance in noisy environments, we use a loudspeaker to play causal human conversations as background noise, and compare the robustness of \ours with replay attack and recording eavesdropping attack.
For the replay attack, we place the attacker (iPhone 8) 30cm away from the victim iPhone X, and replay voice commands with its maximum volume. 
For \ours attack, we use the same experiment setup in Section~\ref{inject_quality}.
The average noise level in quiet environment is 25 dB, and we repeat all the experiments with different background noise levels.
Fig.~\ref{fig:inj_rob} shows the robustness comparison of replay  and \ours injection attacks. 
When the noise level is below 30 dB, both replay attack and \ours achieve 100\% ASR.
However, when the noise level is increased above 45 dB, the ASR of replay attack  drops significantly. In the environments where noise level is above 55 dB, the replay attack cannot succeed.
In contrast, \ours leverages electric signals rather than acoustic signals to inject voice commands. As a result, the external noise will not affect the received audio signals. In all the noisy environments, \ours injection attack can always achieve 100\% ASR, which demonstrates the robustness of \ours injection attack.

\subsubsection{\ours Eavesdropping Robustness}\label{eavesdropping_robustness}
Moreover, we evaluate the robustness of \ours eavesdropping attack on the iPhone X. Similar to the setup in Section~\ref{eavesperfor}, we use an iPhone 8 as the recording device and compare the audio recognition with \ours eavesdropping attack.

\begin{figure}
    \centering
    \includegraphics[width=8cm]{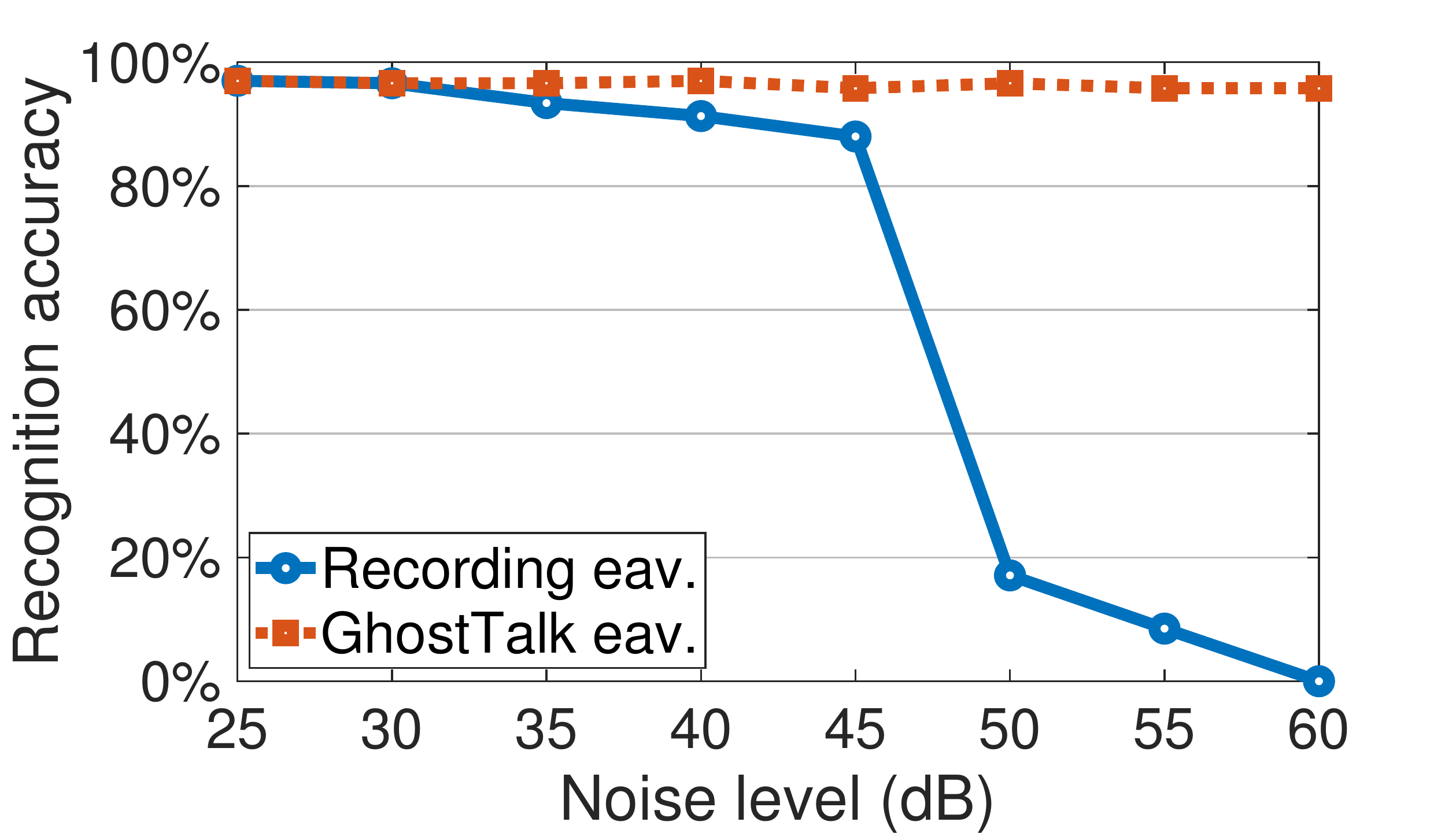}
    \caption{Recognizability comparison of recording eavesdropping and \ours eavesdropping attacks in different noisy environments.}
    \label{fig:eav_rob}
\end{figure}

Fig.~\ref{fig:eav_rob} illustrates the recognition accuracy comparison of recording eavesdropping and \ours eavesdropping attack. 
Unsurprisingly, for normal recording eavesdropping, the recognition rate degrades when the environment noise becomes stronger. Specifically, when the background noise level is higher than 50 dB, Google Speech-to-Text API can hardly recognize the speech contents. 
On the contrary, for the audios recovered by \ours eavesdropping, their perceptibility remains at a constant level. 
Since the external noise has no impact on the electric signals, \ours eavesdropping can still recover clear speech audios in noisy environments.

\subsubsection{\ourss Eavesdropping Robustness}\label{sc_robust}
To evaluate the robustness of \ourss eavesdropping through standard cables, we repeat the experiment in Section~\ref{digitclass} in the environments with different noise levels. We test the robustness on 3 smartphones including iPhone X, Honor 10, and Note 10. All the phones play the testing speech audios with their maximum volume.

\begin{figure}
    \centering
    \includegraphics[width=8cm]{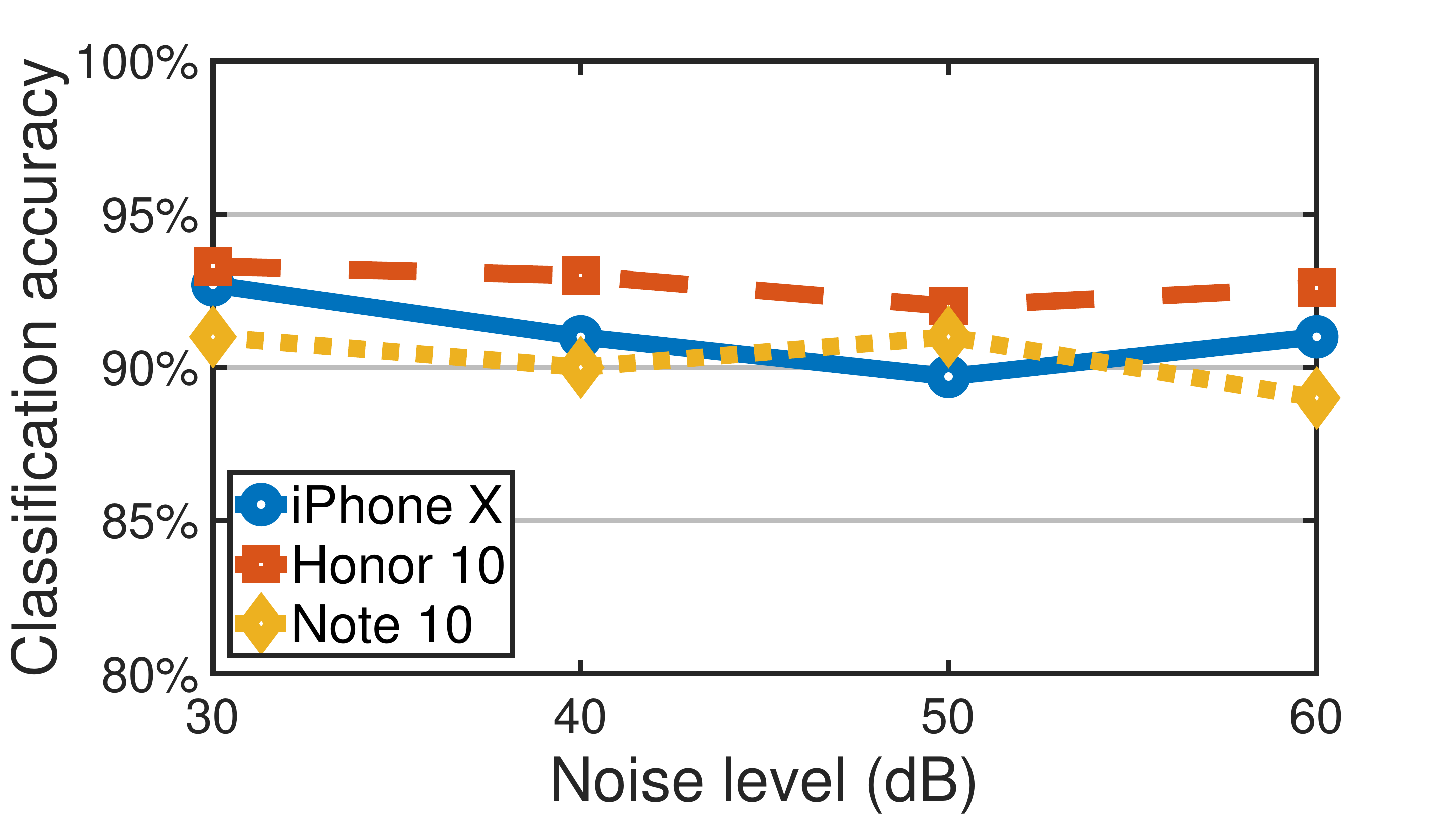}
    \caption{\ourss eavesdropping attack performance in different noisy environments.}
    \label{fig:clas_rob}
\end{figure}

\begin{figure}
    \centering
    \includegraphics[width=8cm]{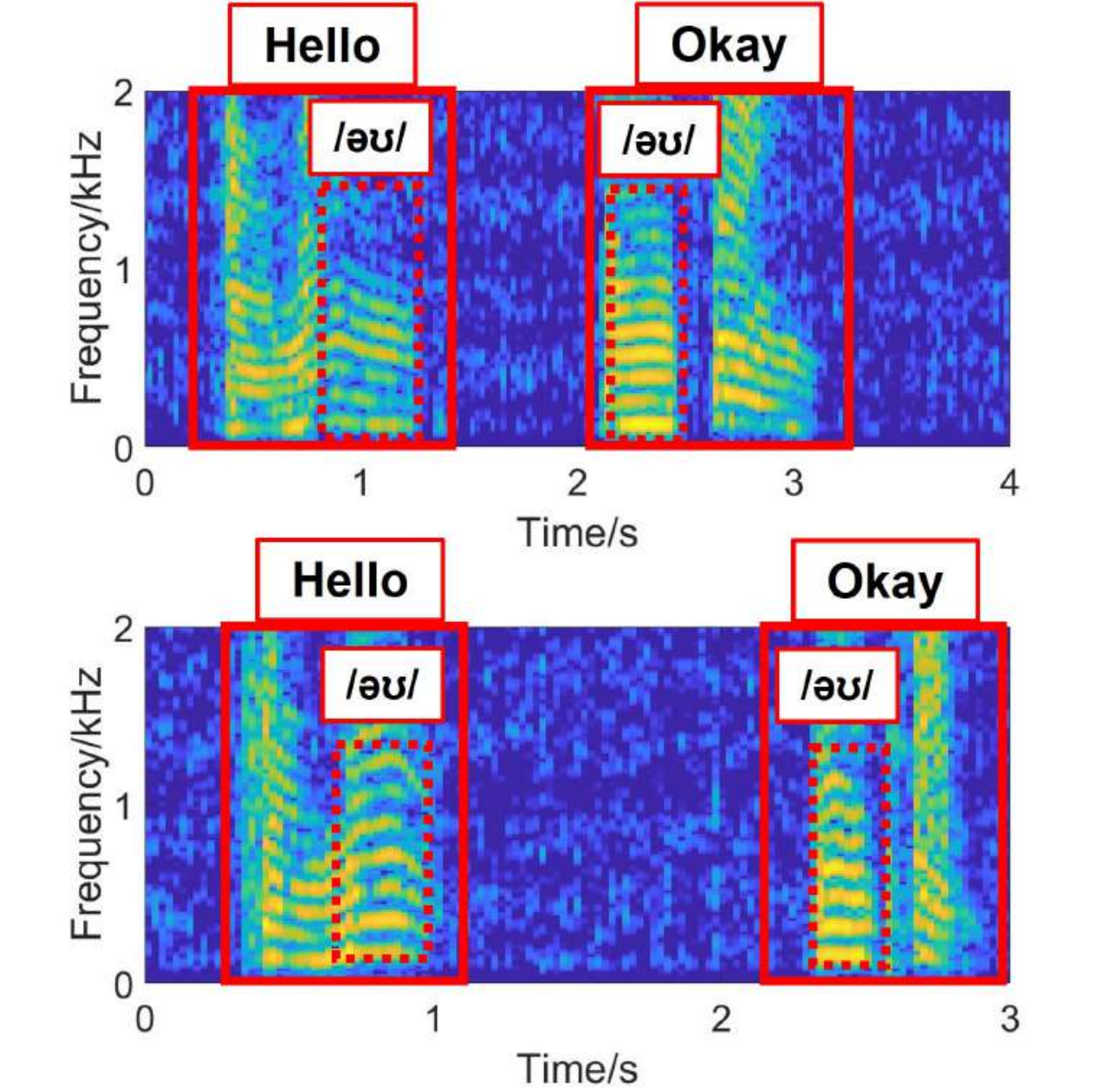}
    \caption{\rev{The spectra of ``hello'' and ``okay" from two volunteers with different speaking speeds.}}
    \label{fig:word}
\end{figure}

Fig.~\ref{fig:clas_rob} presents the robustness evaluation results of \ourss eavesdropping attack. 
We notice that even though the noise becomes increasingly stronger, the digit classification accuracy stays intact. 
The slight difference in accuracy is mainly caused by noise in the current measurement.
The results prove that \ourss eavesdropping attack is robust in noisy environments. It implies that \ourss enables a much wider variety of attack scenarios compared with the existing eavesdropping attacks.
\rev{
\section{Discussion}\label{discussion}
\noindent\textbf{Word Eavesdropping Capability.} In the experiment, we evaluate the digital recognition performance of \ourss. Here, we discuss its potential of word recognition. 
Fig.~\ref{fig:word} shows the spectra of the words ``hello" and ``okay" from two male volunteers with different speaking speeds. Note that 
the denoised spectra of the same word pronounced by different volunteers present similar patterns. Therefore, it is potentially feasible to perform word recognition using GhostTask-SC. However, the CNN model requires a large dataset for the model training. 
Due to the lack of a large dataset containing speech samples of individual words, we cannot verify the word eavesdropping performance. 

Based on the results in Fig.~\ref{fig:word}, we consider two potential approaches that can be used for eavesdropping words. 
First, 
similar to the FSDD dataset for the digit recognition, we can build a common word dataset, which stores spoken word audios from different human speakers. The dataset will be used to train a CNN model for word classification. 
This approach could potentially guarantee a high recognition accuracy, but it requires a large dataset of spoken words. 
Second, Fig.~\ref{fig:word} also shows that the spectrogram components of the same phoneme /\textipa{@U}/ resemble each other regardless of the speaker identities. 
Therefore, we can build a phonetic symbol classification model to recognize phonemes, and the attackers can then recognize the words by annexing the phonemes. 
However, the accuracy of this approach could be degraded due to the difficulty of classifying short phonemes (i.e., they look alike). 
Moreover, the segmentation of phonemes poses a challenge in low-resolution spectra. 
}

\begin{figure}
    \centering
    \includegraphics[width=8cm]{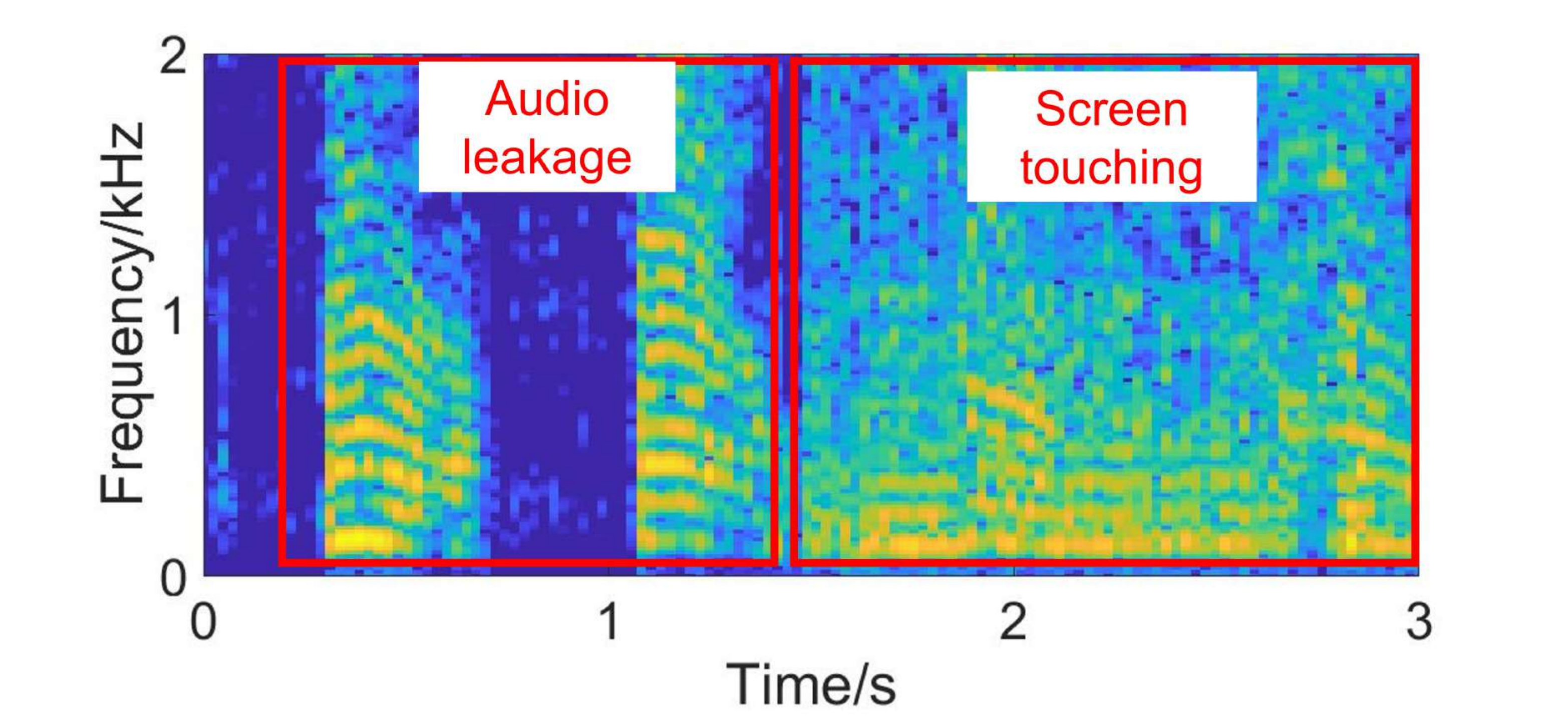}
    \caption{The charging current spectrogram while the user is touching screen and playing audio.}
    \label{fig:touching}
\end{figure}

\noindent\textbf{Touching Screen Interference.} A recent attack, Charger-Surfing~\cite{263834}, demonstrates that the screen touching could produce notable disturbance in the charging current.
Fig.~\ref{fig:touching} shows a charging current spectrogram when the user is touching the screen, and at the same time playing audio with the loudspeaker.
The current noise caused by screen touching is much higher than the idling noise.
Therefore, when the user touches the smartphone screen, the leaked audio signal will be overwhelmed in the strong interference and become unrecognizable. 
But \ourss can still recognize the digits between each screen touching.

\rev{
\noindent\textbf{Attack Stealthiness.} 
The modified cable of \ours in Fig.~\ref{fig:adapter} has the same outlook as a standard cable.
Moreover, our attack system prototype (4.5cm$\times$2.5cm$\times$1cm) is much smaller than a typical 20,000 mAh power bank (16.5cm$\times$7.5cm$\times$2.3cm). We can further shrink the size of the attack system by a better Printed Circuit Board (PCB) design. Therefore, it is realistic to hide the attack system board inside a power bank. 
Without opening up the power bank, the victims will not be able to notice the presence of the attack device. 
As for \ourss attack, it is also quite feasible to hide the attack device behind the USB ports to make the attack stealthy as shown in other studies~\cite{lau2013mactans, keysweeper}.
}


\section{Countermeasure Recommendations}\label{countermeasures}
\noindent \textbf{Disable the Voice Assistant Activation by Headphones.} The key component of the \ours attack is the activation of the voice assistant via shorting the microphone and audio ground wires. If the user disables this option, the attacker will be unable to activate the voice assistant to launch the attack.

\noindent \textbf{Headphone Notification.} Some Android smartphones, such as Huawei Honor 10, will display a ``headphone detected" notification when a headphone is plugged in. 
If the attacker implements the \ours attack on such types of smartphones, the users may be alerted by the notification and realize the underlying threat in the shared power bank.

\noindent \textbf{Stop Charging After Reaching High Percentage Battery Level.} \rev{Note that only the smartphones with their battery level exceeding 95\% can be eavesdropped by \ourss attack.
If the victim stops charging before the battery state reaches that high level, \ourss attack can be effectively avoided.}

\section{Related work}
\subsection{Inaudible Voice Command Injection} 
Inaudible voice command has been a serious threat for voice control systems. Backdoor~\cite{roy2017backdoor} first illustrates the nonlinearity of microphones and transmits audio signal through inaudible ultrasound band. DolphinAttack~\cite{zhang2017dolphinattack} further leverages this nonlinearity to implement a hidden voice command attack to compromise voice control systems. LipRead~\cite{roy2018inaudible} enables a long-range inaudible voice command attack using a speaker array. 
SurfingAttack~\cite{yan2020surfingattack} uses the guided ultrasound wave to inject inaudible voice commands though solid medium. 
However, these attacks must have prior-knowledge about the authorized user's voice, and the attack success rate drops in noisy environments.
Recently, Sugawara et al.~\cite{sugawara2020light} propose a long-range inaudible voice  injection attack by projecting light signals to influence smart device microphones. 
Light commands directly modulate the voice commands on the light signals, making it resilient against noises. 
However, this attack only works in a line of sight scenario.
Compared with existing voice command injection attacks, \ours explores a new backdoor in the smartphone charging port to inject inaudible voice signals.
\subsection{Side-channel Eavesdropping Attacks} 
The loudspeaker of a smartphone vibrates when playing audio signals. The vibration side-channel can be exploited 
to implement the eavesdropping attacks by leveraging the smartphone motion sensors. GyroPhone~\cite{michalevsky2014gyrophone} and Speechless~\cite{anand2018speechless} successfully recognize speaker identity and recover speech contents from motion sensor data. 
AccelEve~\cite{ba2020learning} uses a motion sensor with a higher sampling frequency to further improve the attack performance, and it incorporates a DNN model to recognize the spoken digits. 
However, the attack performance is limited by the sampling frequency. 
Moreover, if the system requires the permissions for the access of motion sensor data, the attacks will no longer succeed.
Recent research has also discovered that it is possible to eavesdrop audio signals by sensing the object vibration. 
ART~\cite{wei2015acoustic} can eavesdrop loudspeakers by measuring the reflected wireless signal strengths and phase differences. 
Lamphone~\cite{nassi2020lamphone} can remotely eavesdrop audio signals by monitoring the slight changes in the brightness of vibrating bulbs. 
LidarPhone~\cite{sami2020spying} exploits Lidar sensors on robot vacuum cleaners to measure the vibration of objects, which can effectively recover the private human speech. 
However, other audio sources may also interfere with the sensing of object vibration, which leads to the degraded attack performance in noisy environments.
Compared with existing work, \ours eavesdropping directly recovers pure audios played by the smartphone, and \ourss can spy private information by extracting leaked audio signals from the power side-channel.
\subsection{Attacks via Charging Cables and Power-Line Channels} 
\rev{Malicious charging cables have been developed to compromise the smartphones. 
Lau et al.~\cite{lau2013mactans} successfully inject malware into iOS devices via malicious chargers.
Shiroma et al.~\cite{shiroma2017threat} successfully spy the victim device screen using a malicious USB cable.
Spolaor et al.~\cite{spolaor2017no} further launch an attack to eavesdrop the sensitive information of Android smartphones from USB cables without requiring any permissions.
In comparison, \ours leverages the audio function backdoor in charging ports and hacks the voice system by deploying malicious charging cables.}  

In addition, the smartphone charging power is influenced by the smartphone apps when the battery state reaches a high level. 
Yang et al.~\cite{yang2016inferring} present an attack that fingerprints user's webpage history by monitoring the power usage, when the user is charging from a public power source. 
Cour et al.~\cite{10.1145/3460120.3484733} further 
extend the fingerprinting attack towards wireless charging devices.
POWERFUL~\cite{chen2017powerful} infers sensitive app usage through smartphone's power consumption profiles. 
Charger-Surfing~\cite{263834} can recover the smartphone's lock screen password by monitoring the charging voltage.
In this work, \ourss utilizes the power side-channel to eavesdrop audios from the charging smartphones. 

\section{Conclusion}
In this paper, we explore the feasibility of voice injection and eavesdropping attacks via the power line. 
With a modified cable, \ours can remotely inject and eavesdrop audio signals through the charging cable, enabling new interactive attack scenarios. \ours does not need any authorized voice information and can work in a noisy environment. Meanwhile, with a standard cable,  we design the \ourss attack to launch an effective audio eavesdropping attack by measuring the charging current through the power line side-channel.
By leveraging a DNN model, \ourss can achieve higher than 92\%  accuracy in identifying spoken digits on various smartphones including iPhone 5s, Honor 10, and MI 8 Lite.

\section*{Acknowledgement}

The authors are grateful to the anonymous reviewers for their constructive comments and suggestions. This work is supported in part by the National Science Foundation grants CNS-1950171 and CNS-1949753. 

Any opinions, findings, conclusions or recommendations
expressed in this material are those of the author(s) and do
not necessarily reflect the views of United States Government
or any agency thereof.



\bibliographystyle{IEEEtran}
\bibliography{reference}

\end{document}